\documentclass[journal]{IEEEtran}
\usepackage{graphicx}
\usepackage{url}
\usepackage{hyperref}
\usepackage{amsmath, amsthm, amssymb, amsfonts} 
\usepackage{siunitx}
\usepackage{dblfloatfix} 
\usepackage{algorithm}
\usepackage{algpseudocode}
\begin{document}

\title{Hardware-Accelerated Geometrical Simulation of Biological and Engineered In-Air Ultrasonic Systems}

\author{
\IEEEauthorblockN{
    Wouter Jansen,
    and Jan Steckel
     }\\
     \IEEEauthorblockA{Cosys-Lab, University of Antwerp, 2000 Antwerp, Belgium}\\
    \IEEEauthorblockA{Flanders Make Strategic Research Centre, 3920 Lommel, Belgium}\\

\thanks{Corresponding author: Wouter Jansen (wouter.jansen@uantwerpen.be)}
}

\maketitle

\begin{abstract}
The deployment of in-air acoustic sensors for industrial monitoring and autonomous robotics has grown significantly, often drawing inspiration from biological echolocation. However, developing and validating these systems in existing simulation frameworks remains challenging due to the computational cost of simulating high-frequency wave propagation in large, dynamic, and complex environments. While wave-based methods offer high accuracy, they scale poorly with frequency and volume. Conversely, existing geometric acoustic solvers often lack support for dynamic scenes, complex diffraction, or closed-loop robotic integration. In this work, we introduce SonoTraceUE, a high-fidelity acoustic simulation framework built as a plugin for Unreal Engine. By using a hardware-accelerated ray tracing-based specular reflection model, and a curvature-based Monte Carlo diffraction model, the system supports real-time simulation of active and passive acoustic sensing in dynamic, multi-material environments. We validate the framework through two distinct experimental domains: a bioacoustic study and a robotics experiment. Our results show that SonoTraceUE reproduces the spatial structure of real-world recordings, and that acoustic odometry and place recognition degrade in the same locations and to a similar degree in simulation as on the real sensor along a common replayed trajectory. The framework provides a versatile platform for synthetic data generation, hypothesis testing in bioacoustics, and the rapid prototyping of robotic systems that use acoustic sensing.
\end{abstract}


\section{INTRODUCTION}
\label{sec:introduction}

The deployment of in-air acoustic sensors has accelerated over the last decade in a variety of industrial and research sectors, driven by their unique sensing capability and versatility in environments where other modalities may fail. On the one hand, these systems are passively used to monitor other acoustic energy sources. Examples include non-contact fault detection in machinery \cite{PACHECOCHERREZ2022106515, alousifMachineryFaultDetection2021, dadoucheSensitivityAirCoupledUltrasound2008, KUNDU2024124169, steckelToolWearPrediction2024}, and the identification of pressurized gas or air leaks in industrial infrastructure \cite{guentherAutomatedDetectionCompressed2016, Wangs18092985, schenck2025}. On the other hand, active acoustic systems, specifically pulse-echo sensors, have seen increased adoption as sensing components in robotics applications. 

\newpage

For autonomous navigation and mapping, active acoustics provide a robust alternative to optical systems such as LiDAR and cameras, particularly in scenarios where visual occlusion caused by dust, fog, smoke, or darkness renders optical data unreliable \cite{moto:c:irua:105050_stec_biom, moto:c:irua:187594_jans_sona, technologies10030054}. The development of these engineered systems frequently draws inspiration from biology. Nature offers highly sophisticated in-air acoustic models, most notably in bat species that rely on echolocation to navigate complex environments and capture prey \cite{schnitzler2003spatial, geipel2013perception, schonerBatsAreAcoustically2015}. Furthermore, experimental and numerical evaluations of engineered airborne ultrasonic arrays, such as Allevato et al.'s characterization of high-precision spiral phased arrays for beamforming and imaging \cite{allevato2022aircoupled}, highlight the importance of understanding array directivity and spatial resolution in hardware design.

Whether investigating these biological mechanisms or engineering their artificial counterparts, simulation and modeling can be an indispensable stage when working with these systems. It allows researchers to model specific sensor configurations \cite{Chen9414446, umadi2025widefield}, test scenarios where data is not (yet) available \cite{machines7020042, schroder2011raven}, generate large synthetic datasets \cite{echoptpaper}, and validate applications in safe and controlled environments before real-world deployment \cite{11169971, chen10802636}. Specifically for biological systems, simulation and modeling allow insight into how some species use sound as input and into how they behave. They can also help take the design and application of bio-inspired human-made sensors to the next level\cite{moto:c:irua:165435_simo_bioi, zhuValidationStudyBatinspired2023, 10.1371/journal.pone.0241443}.

\begin{figure}
    \centering
    \includegraphics[width=0.95\linewidth]{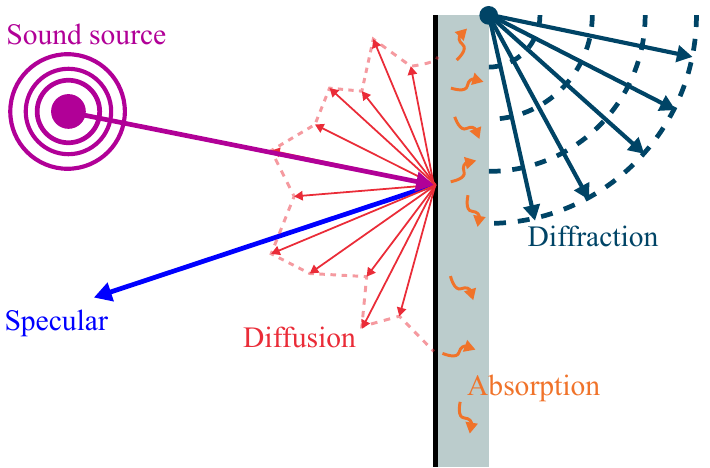}
    \caption{The various acoustic phenomena that are modeled in the abstraction model as proposed in this paper. The big omissions are refraction and more complex forms of medium transmission (e.g., for in-water simulation). Specular, diffuse, and absorption are simulated directly using ray tracing. In contrast, diffraction uses a Monte Carlo approximation to the wave equation solution, based on the local curvature calculated from the scene geometry.}
    \label{fig:acousticphenomena}
\end{figure}

However, simulating acoustic propagation remains computationally challenging, and the choice of technique depends strongly on the problem's scale and frequency. Sound propagation can be divided into several sub-models, including specular and diffuse reflections, absorption, refraction, and diffraction. These are visualized in Figure \ref{fig:acousticphenomena}. Depending on the desired abstraction model, simulation can only tackle some of these phenomena. Wave-based methods, such as the Finite Element Method (FEM) and the Boundary Element Method (BEM), solve the Helmholtz equation directly to describe wave propagation in the frequency domain. These methods provide direct full-wave solutions for the amplitude and phase of the sound pressure at every point in a domain, making them ideal for near-field simulation and detailed analysis of bounded problems. However, this accuracy comes at the cost of computation; as the frequency increases, the required mesh density (geometric resolution) grows drastically.

\newpage 

The computational time required to solve the sparse matrices generated by FEM is typically $\mathcal{O}(N^2)$, depending on the solver. And as BEM produces a dense matrix, it typically requires $\mathcal{O}(N^3)$ operations using standard direct solvers. Consequently, FEM and BEM are often limited to small geometrical scales or low frequencies.

To address scenarios involving larger environments and higher frequencies (ultrasound), as typically seen in echolocation scenarios, geometric acoustics (or ray acoustics) offers a different abstraction model. Unlike the full-wave equation, ray acoustics models sound propagation as rays along which acoustic energy is transported, a technique similar to those used in the optical spectrum. This approach is valid when objects are acoustically large but can be ill-suited when, for example, diffraction effects are required, as they are generally not ideally modeled with geometric acoustics \cite{pierceAcousticsIntroductionIts2019}. In general, ray acoustics allows the simulation of complex, large-scale geometries with mid-to-high-frequency detail that would be impossible or impractical to resolve using wave-based methods. However, the physical dimensions of the scene should be significantly greater than the signal wavelength so that the objects of interest in that scene have high Helmholtz numbers.

In our previous work, we introduced SonoTraceLab\cite{jansenSonoTraceLabAraytracingBasedAcoustic2024}, an open-source MATLAB simulation framework that uses a ray-based acoustic model. It was purpose-built to fill a gap left by commercial and existing open-source solutions, specifically for 3D ensonification studies, synthetic data generation, and closed-loop testing of biological hypotheses within the frequency range of \SI{20}{\kilo\hertz} to \SI{200}{\kilo\hertz}. SonoTraceLab demonstrated effectiveness in providing insights into biological echolocation while reducing the complexity of physical experiments. 

Nevertheless, it faced some architectural limitations. The environment relied on static 3D meshes with a single globally defined surface material, modeled using the Bidirectional Reflectance Distribution Function (BRDF). 

\newpage 

These limitations restricted the ability to model complex, dynamic, multi-material scenes. Furthermore, while the core ray tracing was GPU-accelerated, reliance on MATLAB introduced latency, limiting performance. It also lacked support for multiple simultaneous acoustic sources and was not optimized for rapid scenario creation, as it lacked a 3D editor interface.

\begin{figure}[!b]
    \centering
    \includegraphics[width=1\linewidth]{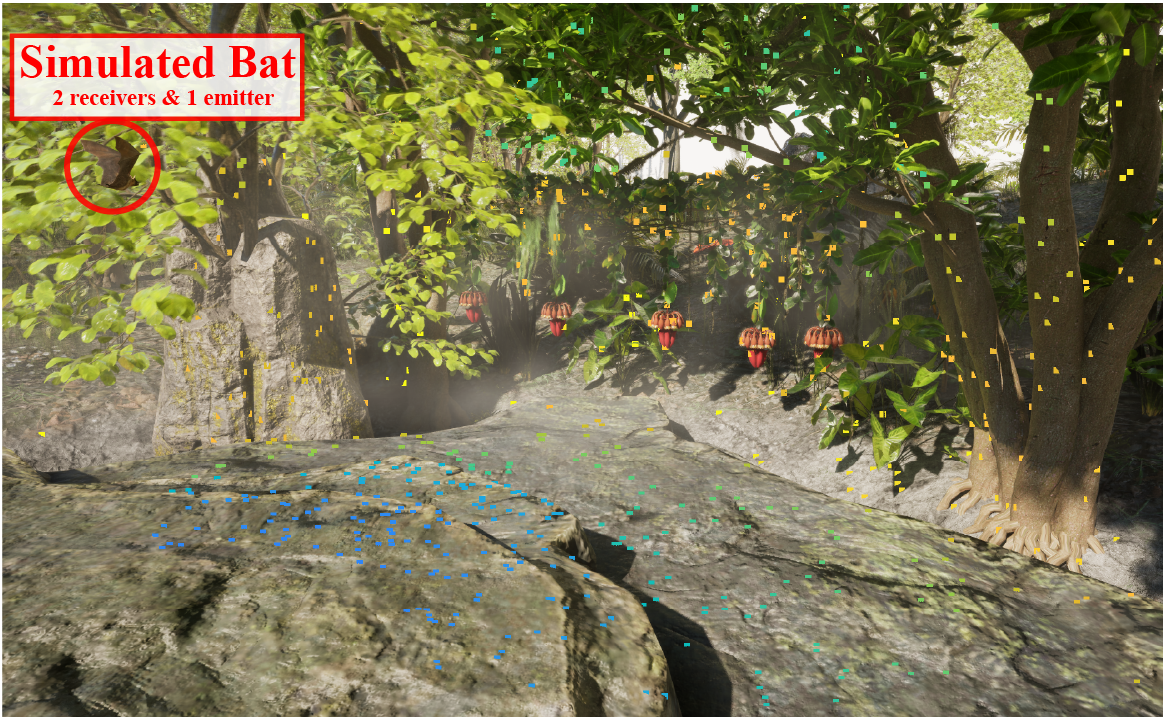}
    \caption{An example of a large forest environment simulated with SonoTraceUE in Unreal Engine, consisting of 96 unique 3D meshes totaling over \SI{10000} instances for a combined \SI{21000000} triangles. The colored points shown in the image are representative of a small portion of the simulated points by the simulation of a bat, which has two receivers and one emitter in this scenario. A video of the simulation running in this scene is available as supplementary material and online \cite{sonotraceueyoutube}.}
    \label{fig:scene}
\end{figure}

In this paper, we introduce SonoTraceUE, a novel implementation that overcomes these scalability and usability constraints and that is built upon the Unreal Engine framework \cite{unrealengine}. As a multi-platform real-time 3D rendering engine, Unreal offers high-fidelity visualization and optical rendering, along with a versatile editor for creating and scripting dynamic scenarios across large environments. SonoTraceUE leverages this architecture as a plugin, enabling a user-friendly workflow for simulating acoustic systems in large-scale, high-fidelity, and dynamic scenes, as shown in the example in Figure \ref{fig:scene}. By leveraging hardware-accelerated ray tracing, the framework enables simulation at rates suited to bioacoustic hypothesis testing and offline generation of robotic algorithm-development data. It can simulate multiple active or passive acoustic systems interacting with fully modeled BRDF surface properties.

We situate our work within the state of the art in acoustic system simulation and detail the methodological advancements of SonoTraceUE over its predecessor. We present a series of practical robotic experiments comparing simulated and real-world acoustic odometry and localization along a common, replayed trajectory to validate the framework's fidelity. Beyond validation, we demonstrate its capabilities in scenarios impractical to replicate physically, thereby deepening our understanding of biological echolocation in a practical bioacoustic experiment. We analyze the computational trade-offs of geometric simulation at various resolutions and conclude with our vision for the future trajectory of geometric acoustic simulation.

\section{RELATED WORKS}
\label{sec:relatedworks}

The landscape of acoustic simulation is vast, with methodologies and simulation frameworks often purpose-built for specific frequency ranges, domains, mediums, or applications. As acoustic wave propagation is well understood \cite{pierceAcousticsIntroductionIts2019}, the level of abstraction required determines which simulation model fits best. Broadly, these can be categorized into wave-based numerical methods, geometric acoustics, and domain-specific solvers for biomedical or robotic applications.

\subsection{WAVE-BASED SIMULATION}
\label{subsec:wavebasedsimulation}
Wave-based and multiphysics solvers are often used for high-fidelity analysis of wave phenomena, such as diffraction, interference, and resonance in complex media. FEM\cite{ihlenburgFiniteElementAnalysis1998, thompsonReviewFiniteelementMethods2006} and BEM\cite{kirkupBoundaryElementMethod2019} are the industry standard approaches. Several commercial multiphysics platforms offer robust solvers such as COMSOL \cite{comsolcomsolCOMSOLMultiphysicsSoftware}, ANSYS\cite{ansyssoundsim}, and Siemens SimCenter 3D \cite{siemensSiemensNXAcoustics}. These tools excel in analyzing bounded problems, for example, for automotive cabin acoustic analysis or component vibration. Similarly, k-Wave, an open-source MATLAB framework, uses a k-space pseudo-spectral method to simulate wave propagation \cite{treebyKWaveMATLABToolbox2010}. Field II is also a widely cited reference for simulating the spatial impulse response of linear ultrasonic transducers \cite{field-iiFieldIIUltrasound}. These wave-based approaches have a high computational cost but do offer high simulation accuracy. As the simulation frequency increases to the ultrasonic range, the required mesh density becomes prohibitively expensive for large-scale, three-dimensional environments, rendering them less suitable for simulating extensive robotic navigation tasks. Methods based on FEM generally scale cubically with voxel size in 3D, which can become computationally problematic for large dynamic scenes \cite{caoParallelNumericalAcoustic2020}.

\subsection{GEOMETRIC ACOUSTICS}
\label{subsec:geometricacoustics}
To address the computational scaling issues of wave-based methods, geometric acoustics approximates sound propagation in a different way. Within this domain, two primary techniques exist: the Image-Source Method (ISM) and ray tracing. ISM geometrically mirrors the sound source across reflective boundaries to calculate the exact path of specular reflections \cite{10.1121/1.382599}. While highly accurate for early reflections, its computational cost grows exponentially with the order of reflections. Ray tracing employs a sampling approach, casting discrete rays into the environment to estimate acoustic energy propagation \cite{onakaDesign3dimensionalSound2009, roberHRTFSimulationsAcoustic2006, roberRayAcousticsUsing2007}. This stochastic approach makes ray tracing significantly more efficient for modeling complex geometries and late reverberation.
Commercial packages such as Odeon \cite{odeon} and EASE \cite{ease} combine image-source methods with ray tracing to predict parameters like reverberation time and speech intelligibility. The previously mentioned Siemens SimCenter 3D also includes a geometric acoustic option \cite{SiemensRayAcoustics}. Research-oriented tools also exist, such as RAVEN, developed by Schröder et al., which offers real-time binaural auralization \cite{schroder2011raven}. The MATLAB Audio Toolbox also includes a stochastic ray tracer for room impulse response generation \cite{audiotoolbox}. 

Additionally, the web-based CRAM tool has provided an educational platform for understanding these phenomena\cite{cram}. While these systems are highly validated for audible frequencies in static environments, they often lack support for the larger dynamic scene changes required in robotics without computationally expensive pre-calculation phases. Furthermore, integrating these standalone applications into a closed-loop robotic control stack can be challenging. 

\begin{figure*}[!t]
    \centering
    \includegraphics[width=1\linewidth]{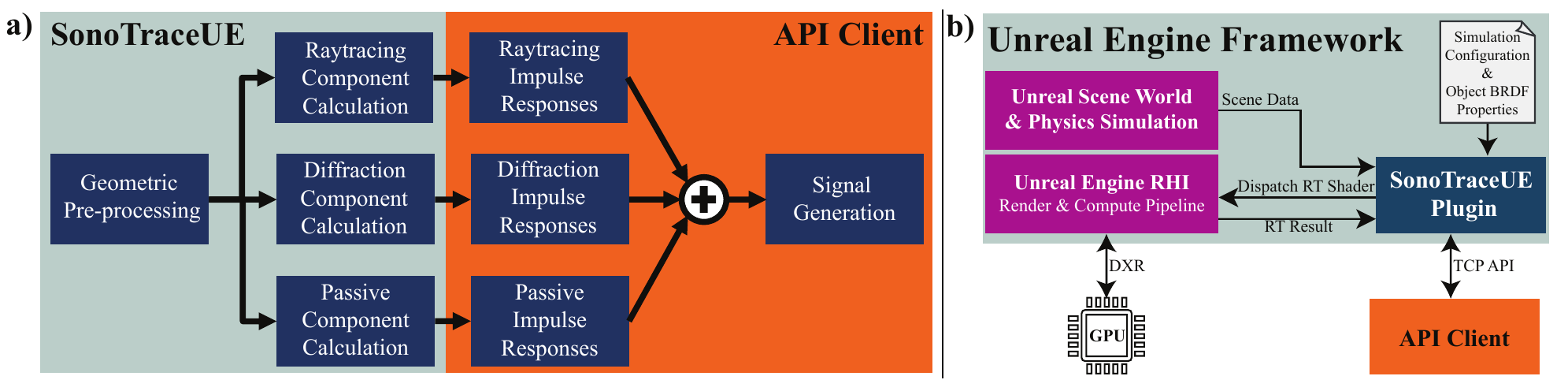}
    \caption{Diagrams visualizing the architecture of the simulation framework. a) Block diagram showing the sequential stages of the simulation implementation. It shows the three main components: specular, diffraction, and passive sensing, respectively. The impulse response is generated, and the receiver signals are created in third-party software via the TCP API. b) High-level schematic of the proposed system architecture integrated within the Unreal Engine framework. The plugin serves as the main block controlling and executing the simulation pipeline, acting as an intermediary between the host application's CPU thread, which handles rigid-body physics and dynamic scene updates, and the low-level render dependency graph that uses the Rendering Hardware Interface (RHI) of Unreal Engine to access the GPU. Acoustic propagation is offloaded to the GPU via asynchronous compute shaders utilizing DirectX Ray Tracing (DXR) pipelines. Interaction with the simulation and final signal post-processing is available through a TCP API to third-party software, such as MATLAB or Python.}
    \label{fig:combined_diagrams}
\end{figure*}

Geometric acoustics shares a fundamental mathematical lineage with optical ray tracing to improve computational efficiency \cite{liu2021soundsynthesispropagationrendering}. Modern optical rendering relies on path tracing, which solves the rendering equation by stochastically tracing millions of light paths. Recently, researchers have begun adapting these mature optical frameworks for acoustics. For instance, Cao et al. adapted bidirectional path tracing to solve the acoustic rendering equation \cite{10.1145/2980179.2982431}. At the same time, Schissler et al. leveraged the NVIDIA OptiX graphics framework to achieve real-time simulation of high-order diffraction \cite{10.1145/2601097.2601216}. Finally, Finnendahl et al. used the differentiable rendering engine Mitsuba 3 to implement differentiable acoustic path tracing \cite{10.1145/3730900}, allowing the simulation of inverse problems such as recovering scene parameters from observations. 

\subsection{STOCHASTIC AND STATISTICAL MODELING}
\label{subsec:stochasticmodeling}
For complex volume scatterers such as foliage or naturally occurring surfaces, explicit wave-based simulation or geometric reconstruction can be computationally expensive. In these scenarios, acoustic returns are sometimes approximated using stochastic signal modeling. These methods treat the target as a random process, synthesizing echoes based on statistical properties such as the power spectral density, envelope statistics, or autocorrelation \cite{yovel2008, grunwald2004classification, kuc2020artificial}. While these statistical approaches are highly effective for texture classification and modeling macroscopic clutter envelopes, they inherently randomize the microscopic temporal structure of the echo. Our proposed framework explicitly avoids stochastic approximation in favor of the deterministic temporal fidelity provided by geometric acoustics.

\subsection{DEEP LEARNING}
\label{subsec:deeplearning}
Recently, machine learning has emerged as a novel paradigm across the acoustic simulation pipeline. As comprehensively reviewed by Liu and Manocha \cite{liu2021soundsynthesispropagationrendering}, deep learning architectures are increasingly employed to map 3D object features to mode data for impact sound synthesis \cite{jin2020deepmodal} or to synthesize musical instruments \cite{hawley2020synthesis, ji2020comprehensive}. In sound propagation and rendering, convolutional and field-based neural networks have been utilized to approximate acoustic scattering \cite{fan2020fast, tang2020learning}, estimate filter parameters \cite{pulkki2019machine}, and derive room material properties from audio recordings or images \cite{schissler2017acoustic, tang2020scene}.

Despite these advancements, the primary advantage of the proposed deterministic geometric approach over current deep learning methods lies in physical explainability and temporal fidelity. 

\newpage 

Deep learning models often approximate sound propagation through learned latent representations, which can statistically smooth over critical microscopic temporal structures and multi-path interference patterns. Furthermore, these data-driven models and their training datasets are almost exclusively constrained to the audible frequency spectrum. They lack the spatial resolution required to accurately model the complex scattering and highly directional wave phenomena characteristic of ultrasound. In contrast, SonoTraceUE calculates discrete, deterministic reflection and diffraction paths based on explicit physical formulas. For specialized tasks such as simulating the high-frequency ultrasonic pulse-echo systems used in bio-inspired echolocation and robotic mapping, preserving this exact temporal structure—including precise time of arrival and phase differences across a microphone array—is essential. The proposed geometric framework ensures that these crucial signal components are explicitly resolved.

\subsection{ROBOTICS SIMULATION}
\label{subsec:roboticssimulation}
In the field of robotics, high-fidelity visual and physical simulation is well established through platforms such as Gazebo \cite{Koenig2004DesignSimulator}, NVIDIA Isaac Sim \cite{NVIDIA_Isaac_Sim}, and CARLA \cite{Dosovitskiy2017CARLA:Simulator}. These simulators excel at rigid-body dynamics, photorealistic rendering, and the simulation of optical sensors such as LiDAR and cameras. However, their acoustic simulation capabilities are often limited. Acoustic sensors are typically modeled as idealized ray casts or simple object-collision interaction calculations, lacking the complex multi-path propagation, diffraction, and frequency-dependent absorption required for realistic sonar or acoustic scene analysis. Cosys-AirSim, which, like our proposed solution, leverages Unreal Engine to introduce more advanced sensor models, includes a CPU-based geometric wave simulation implementation for active and passive sensors such as radar and sonar\cite{cosysairsim2023jansen, 11267032}, but lacks complex per-surface BRDF modeling, support for dynamic scenes for passive sensing, and arbitrary multi-receiver/multi-emitter array configurations. 

\subsection{SONOTRACELAB}
\label{subsec:sonotracelab}
Our previous open-source framework, SonoTraceLab\cite{jansenSonoTraceLabAraytracingBasedAcoustic2024}, aimed to bridge the gap between biological studies and engineering simulation by using a MATLAB-based ray-acoustic solver. It provided a tool for 3D ensonification experiments and synthetic data generation for medium-sized static scenes. It supported defining surface material properties using a BRDF. It used a GPU-accelerated ray tracing implementation running on NVIDIA CUDA for fast calculation of the specular reflections. It could simulate a single sound source and multiple receivers. However, reliance on the MATLAB interpretation layer prevented true high performance in larger environments, and the system was constrained to static meshes with a single definition of material properties. Using MATLAB also does not offer a user-friendly interface for scenario creation. Our proposed approach follows and implements the same validated and demonstrated simulation technique as in  SonoTraceLab, with additional improvements and features, as well as changes to the technical implementation. 

\section{METHODS}
\label{sec:methods}
This proposed framework is available as open-source and natively supports array configurations of emitters (sources) $s$ and receivers (microphones) $m$. We will cover all the implementation details in this paper for completeness. Note that some mathematical symbols are different from those in the original SonoTraceLab manuscript. 

\subsection{ARCHITECTURE}
\label{subsec:Architecture}
Figure \ref{fig:acousticphenomena} shows the specific acoustic phenomena our abstraction model handles. As can be seen, refraction and more complex wave transmission are not modeled. Figure \ref{fig:combined_diagrams}a shows the flowchart of the simulation. The three main components are shown as specular, diffraction, and passive, respectively. 

The passive component calculates the direct line of sight between receivers and sound sources (emitters) and is primarily used when simulating passive sensing systems.  This figure also shows a clear split between the steps calculated within the Unreal Engine plugin SonoTraceUE and those performed in the external API Client. Currently, we have created one such API Client in MATLAB.

The architecture is further visualized in \ref{fig:combined_diagrams}b. The core simulation module, SonoTraceUE, is an Unreal Engine plugin. Unreal is a multi-platform real-time 3D rendering engine, allowing the usage of hardware-accelerated ray tracing using the Unreal Engine Rendering Hardware Interface (RHI). It also has a user-friendly and versatile editor that supports C++ \& node-based visual scripting, allowing the user to create large and dynamic acoustic simulation scenarios with support for closed-loop interactions. The plugin retrieves all data required for the scene, including the transformations of all objects and their mesh data. The interaction with the API Client is done over an API interface.  

\subsection{GEOMETRIC PRE-PROCESSING}
\label{subsec:pre-processing}
In a first offline pre-computation step, all registered objects in the scene are retrieved, and the 3D mesh data is loaded into CPU memory. SonoTraceUE supports both static and skeletal meshes, with the latter capable of animation. During this step, an offline analysis of the scene geometry is performed to identify acoustically significant features, such as edges and holes. As a metric of acoustic significance, a local sharpness or curvature metric is calculated for each $i$-th triangle $T_i$ of the mesh. First, the discrete mean curvature normal $\mathbf{K}_v$ is calculated for every vertex $v$ utilizing the discrete differential geometry operators defined by Meyer et al. \cite{meyercurvature2003}. The scalar mean curvature magnitude is then given by $G_v = \frac{1}{2}\|\mathbf{K}_v\|$.

To robustly detect geometric discontinuities while ignoring smooth surface undulations, we define the curvature variation $\Delta G_i$ of the $i$-th triangle as the range of curvature magnitudes among its adjacent vertices. To ensure consistency across varying mesh resolutions, this metric is modulated by a piecewise-linear area weighting function $w(A_i)$ and a global scaling factor $\eta$. The final diffraction curvature metric $C_i$ is defined as:

\begin{equation}
\label{eq:curvature}
C_i = \eta \cdot w(A_i) \cdot \underbrace{\left( \max_{v \in T_i} G_v - \min_{v \in T_i} G_v \right)}_{\text{Curvature variation } \Delta G_i} \end{equation}

Here $A_i$ is the area of triangle $T_i$. $w(\cdot)$ is a piecewise-linear function of triangle area, set by three breakpoints $A_1 < A_2 < A_3$. It starts at $w_{\min}$ for very small triangles, rises to unity at $A_1$, continues up to a peak $w_{\max}$ at $A_2$, and then falls back to zero at $A_3$, beyond which triangles no longer contribute to $C_i$. $\eta$ is a user-set scaling constant. As originally proposed and validated in \cite{jansenSonoTraceLabAraytracingBasedAcoustic2024}, this metric $C_i$ effectively acts as a high-pass geometric filter, responding strongly to sharp edges necessary for diffraction modeling while vanishing on planar or smoothly curved surfaces. 

\newpage 
From this metric, we compute two frequency-dependent surface material properties, $\beta(T_i,f)$ and $k(T_i,f)$, for each triangle: the opening angle and the reflection magnitude of the acoustic BRDF, respectively. $\beta$, in degrees, governs the angular distribution of reflected energy and effectively determines the transition between specular and diffuse scattering. At the same time, $k$, a dimensionless pressure-amplitude reflection factor in $[0,1]$, represents the frequency-dependent reflection coefficient and scales the magnitude of the signal returned from the triangle. Concretely, both are obtained from $C_i$ by a per-object, per-frequency-bin sigmoidal blend between a smooth and a rough endpoint. Writing the sigmoid function as:
\begin{equation}
\label{eq:sigmoid}
S(C;\sigma,\mu) = \frac{1}{1+e^{-\,\lambda\sigma\,(C-\mu)}}
\end{equation}
The two surface properties can then be defined as:
\begin{equation}
\label{eq:brdfmap}
\begin{aligned}
\beta(T_i,f) &= S_\beta\,\varphi_{1}(f) + (1-S_\beta)\,\varphi_{0}(f)\\
k(T_i,f) &= S_k\,\rho_{1}(f) + (1-S_k)\,\rho_{0}(f)
\end{aligned}
\end{equation}
with $S_\beta = S(C_i;\sigma_\beta,\mu_\beta)$ and $S_k = S(C_i;\sigma_k,\mu_k)$. $\varphi_0,\varphi_1$ and $\rho_0,\rho_1$ are the smooth- and rough-surface end material property variables assigned to the object, linearly interpolated across the simulated frequency band, and $\sigma,\mu$ the slope and center of the transition, set independently for each property.
The constant $\lambda$, set to 8 in all our experiments, is a fixed steepening factor, introduced because a unit slope produces a material transition that is too gradual to be useful in practice. This keeps the assigned material properties bounded by the two end points for any value of $C_i$.

To enhance simulation fidelity, the proposed architecture departs from the single-material constraint of prior work. While SonoTraceLab utilized a monolithic material definition for the entire scene mesh, SonoTraceUE supports per-object material granularity. Each geometric primitive in the scene can be linked to a specific acoustic profile. 

\subsection{RAY TRACING COMPONENT}
\label{subsec:ray tracing}
To solve the Helmholtz equation in our abstraction model, we separate some phenomena, as visualized in Figure \ref{fig:acousticphenomena}. For solving the specular reflections, we use ray tracing. A specified number of rays is emitted into the environment from each emitter source. The rays are distributed using recursive zonal equal area sphere partitioning \cite{leopardiPartitionUnitSphere2006}. For each ray, we calculate the full propagation path. At intersections with an object's surface, we compute the specular reflection around the surface normal of the triangle $T_i$ that was hit on that mesh. Propagation and specular reflections will continue until either the maximum number of bounces or the maximum propagation length is reached. 
The ray tracing algorithm is implemented as a compute shader and is dispatched to the Unreal RHI. It uses the DirectX Ray Tracing (DXR) pipeline to perform ray-triangle intersection calculations on ray tracing cores available on modern GPU devices. Algorithm \ref{alg:ray tracing_shader} shows a simplified version of the ray tracing algorithm. Afterward, the ray tracing result is parsed further in the SonoTraceUE plugin on the CPU. 
\newpage 

\begin{algorithm}
\caption{Acoustic ray tracing Logic Shader}
\label{alg:ray tracing_shader}
\textbf{Input} Source Pose $P_{source}$\\
\hspace*{8.5mm} Ray Angles $\Theta$\\
\hspace*{8.5mm} Scene Geometry $M$ (TLAS/BLAS) \\
\textbf{Output} Hit Data Buffer $B$ 
\begin{algorithmic}
\For{\textbf{each} Ray Angle $\theta_i \in \Theta$}
    \State $\vec{p}_{origin} \gets P_{source}.\text{Location}$
    \State $\vec{d}_{ray} \gets \text{CalculateWorldDirection}(P_{source}.\text{Rotation}, \theta_i)$
    \State $r_{total} \gets 0$
    \For{$j \gets 0$ \textbf{to} $MaxBounces$}
        \State $H \gets \text{DXR.GenerateRay}(\vec{p}_{origin}, \vec{d}_{ray}, M)$
        
        \If{$H.\text{isHit}$}
            \State $\vec{d}_{reflect} \gets \text{CalculateReflection}(\vec{d}_{ray}, H.\text{Normal})$            
            \If{$j = 0$}
                \State $r_{total} \gets \text{Distance}(P_{source}, H.\text{Location})$
            \Else
                \State $r_{total} \gets r_{total} + H.\text{Distance}$
            \EndIf
            
            \State $\vec{v}_{hit} \gets \text{Direction}(H.\text{Location}, \vec{p}_{origin})$
            \State $L_{hit} \gets \text{DXR.GenerateRay}(H.\text{Location}, \vec{v}_{hit}, M)$
            
            \State $B[i][j] \gets \{H, r_{total}, \vec{d}_{reflect}, L_{hit}\}$
            
            \State $\vec{p}_{origin} \gets H.\text{Location}$
            \State $\vec{d}_{ray} \gets \vec{d}_{reflect}$
            
            \If{$r_{total} \geq MaxDistance$}
                \State \textbf{break}
            \EndIf
        \Else
            \State \textbf{break}
        \EndIf
    \EndFor
\EndFor
\end{algorithmic}
\end{algorithm}

For each $n$-th reflection point, a total signal magnitude or strength $M_{rt}(f,n,s,m)$ is calculated over the frequency band $f$ between the sound source or emitter $s$ and receiver (microphone) $m$. In the following equations, we will omit $(n,s,m)$ for simplicity.

For each receiver $m$, the total propagation path length $r$ is calculated in meters from the emitter $s$ along this $n$-th reflection point. With this, we can define the geometric spreading term along the specular reflection path as $L_{s, geo} = \frac{1}{r^{2}}$. This approximates the two-way spreading, more specifically the product of the source-to-reflection-point and reflection-point-to-receiver distances, divided by the square of their sum. This approximation holds well whenever the emitter and receiver are close together relative to the distance to the reflection point.

Atmospheric attenuation $L_{atm}(f)$ is calculated as a function of frequency and total path distance:
\begin{equation}
\label{eq:atm}
L_{atm}(f) = 10^{-\alpha(f)\, r / 20}
\end{equation}
where the absorption coefficient $\alpha(f)$, expressed in \si{\decibel\per\meter}, is a linear fit to the frequency-dependent absorption of air:
\begin{equation}
\label{eq:alpha}
\alpha(f) = a\,\frac{f}{\SI{1}{\kilo\hertz}} + b, \qquad a = 0.038, \quad b = -0.3
\end{equation}
valid for the ultrasonic band targeted by this work.

We define the deviation angle $\gamma$, in degrees, as the angle between the ideal specular reflection vector and the actual direction vector to the receiver from the reflection point. Together with the frequency-dependent surface material properties of that triangle, we use that angle to calculate the specular reflection factor $R_{rt}$, a dimensionless pressure-amplitude factor, for a given frequency band $f$ as:
\begin{equation}
\label{eq:Rrt}
R_{rt}(f) = \exp\left({-\frac{\gamma^2}{2{\beta(T_i,f)}^2}}\right) \cdot k(T_i,f)
\end{equation}

The magnitude $M_{rt}(f)$, a dimensionless pressure amplitude relative to the source, can now be defined as:
\begin{equation}
\label{eq:Mrt}
M_{rt}(f) = L_{s, geo} \cdot L_{atm}(f)\cdot R_{rt}(f)
\end{equation}
No normalization with respect to ray count is applied. Absolute levels are therefore only meaningful within a fixed sampling configuration.

It is this magnitude that is calculated for each reflection point on the CPU, parallelized across the number of points, and stored in a large point cloud data structure within the framework. As described in the architecture, a third-party client performs the next steps after transferring the data via the API. With this point cloud data, which contains all magnitude values for the reflection points, we can create the complete transfer function that encodes the specular, absorption, and diffusion acoustic phenomena in our simulation model. This transfer function $H(f)$ is calculated as follows:
\begin{equation}
\label{eq:Hf}
    H(f) = M_{rt}(f) \cdot e^{j \cdot \frac{\omega}{c} \cdot r}
\end{equation}
with the second term the signal delay from path propagation, using the $e^{+j\omega t}$ Fourier convention under which a delay $\tau$ maps onto $e^{+j\omega\tau}$, and $c$ the speed of sound set to \SI{343}{\meter\per\second}. This complex transfer function can now be transformed into the time domain using an inverse fast Fourier transform $\mathcal{F}^{-1}$:
\begin{equation}
\label{eq:ifftH}
h(t,n,s,m) = \mathcal{F}^{-1} \bigg[ H(f,n,s,m) \bigg]
\end{equation}
For calculating the final impulse response $h_{s,m}(t)$ for the $m$-th receiver and coming from emitter $s$, one has to sum up the impulse responses for the $N$ reflection points in the scene as calculated by the ray tracing shader:
\begin{equation}
\label{eq:hsum}
h_{s,m}(t) = \sum^{N}_{n=1} h(t,n,s,m)
\end{equation}

\subsection{DIFFRACTION COMPONENT}
\label{subsec:diffraction}
The ray tracing component is only valid for larger Helmholtz numbers. Diffraction would require a large number of rays, which would be computationally unfeasible. To still have a fast solution for modeling this component, we proposed in SonoTraceLab a Monte Carlo approximation based on the local curvature values $C_i$ calculated during the pre-compute step to identify a fixed number of high-local-curvature diffraction echo candidates. The same technique is applied here. 

We sort all triangles $T_i$ by curvature and use importance sampling to distribute a fixed number $O_m$ of diffraction candidates per mesh across its geometry, weighted toward the most strongly curved triangles \cite{corsiniEfficientFlexibleSampling2012}. Only meshes that are within the field of view or frustum of the emitter source $s$, as defined in the configuration of the simulation, are candidates for diffraction.
Next, the candidates are filtered. A line-of-sight test rejects candidates occluded from the sensor, and we reject candidates on triangles larger than the area threshold $A_3$, since diffraction is not modeled on large, flat facets. Finally, an orientation test retains only candidates whose surface normal makes an angle greater than \SI{90}{\degree} with the vector from the source to the candidate. Afterward, we follow a similar approach as in the previous section to calculate the magnitude $M_{d}(f)$ for the $o$-th diffraction points of the remaining candidates $O$. First, we calculate the path $r$ between the source $s$, receiver $m$, and the $o$-th diffraction point. The magnitude $M_{d}(f)$ can now be defined as:
\begin{equation}
\label{eq:Md}
M_{d}(f) = L_{d,geo} \cdot L_{atm}(f)\cdot k_d(f)
\end{equation}
The geometric spreading term $L_{d, geo}$ along the path is defined as $L_{d, geo} = \frac{1}{r^{2}}$, similar as for the specular component. 

This method models the diffraction point as a secondary omnidirectional point source, rather than a direction-dependent reflection. The diffraction component is simulated by parallel gathering of diffraction candidate points, enabling concurrent multi-core execution that substantially reduces computation time, even for large, high-resolution scenes with many candidates, while remaining optimized compared to SonoTraceLab. No importance-sampling weight is applied, so the synthesized diffraction level scales with $O$, which must therefore be held fixed across simulations compared in amplitude. 

The synthesis of the transfer function $G(f)$ in the third-party client can now be done similarly to that of the specular component:
\begin{equation}
\label{eq:Gf}
    G(f) = M_{d}(f) \cdot e^{j \cdot \frac{\omega}{c} \cdot r}
\end{equation}
This complex transfer function can now be transformed into the time domain using an inverse fast Fourier transform $\mathcal{F}^{-1}$:
\begin{equation}
\label{eq:ifftG}
g(t,o,s,m) = \mathcal{F}^{-1} \bigg[ G(f,o,s,m) \bigg]
\end{equation}
The final impulse response $g_{s,m}(t)$ for the $m$-th receiver and coming from emitter $s$ requires us to sum the impulse responses:
\begin{equation}
\label{eq:gsum}
g_{s,m}(t) = \sum^{O}_{o=1} g(t,o,s,m)
\end{equation}

\subsection{PASSIVE COMPONENT}
\label{subsec:Passive}
A new optional component has been added to the proposed method, which calculates direct propagation between the source and receiver, which can be useful for simulating passive systems, a feature not available in the original SonoTraceLab. This feature is implemented by performing a line-of-sight test between each source $s$ and receiver $m$, and calculating the propagation distance $r$ between them. 

The configuration scaler parameter $A_{p}$ defines the source pressure amplitude and is a single frequency-independent constant. $M_{p}(f)$ can now be defined as:
\begin{equation}
\label{eq:Mp}
M_{p}(f) = L_{p, geo} \cdot L_{atm}(f)\cdot A_{p}
\end{equation}
As this path is one-way, for this component the geometric spreading loss is defined as $L_{p, geo}=\frac{1}{r}$.

Then we can synthesize the transfer function $P(f)$ as follows:
\begin{equation}
\label{eq:Pf}
    P(f) = M_{p}(f) \cdot e^{j \cdot \frac{\omega}{c} \cdot r}
\end{equation}
This complex transfer function can now be transformed into the time domain using an inverse fast Fourier transform $\mathcal{F}^{-1}$ to get the final impulse response :
\begin{equation}
\label{eq:ifftP}
p_{s,m}(t) = \mathcal{F}^{-1} \bigg[ P(f,s,m) \bigg]
\end{equation}

\subsection{SIGNAL GENERATION}
\label{subsec:signals}
With all partial components calculated, we can generate the scene's total spatial impulse response $h_{\text{tot},s,m}(t)$ for the $m$-th receiver and emitter $s$ by summing up the partial solutions:
\begin{equation}
h_{\text{tot},s,m}(t) = h_{s,m}(t) + g_{s,m}(t) + p_{s,m}(t)
\end{equation}

All frequency-dependent terms in the preceding subsections are evaluated on a discrete set of $F$ bins rather than on the full grid, since the pre-processing cost and the memory footprint of the per-triangle BRDF data scale linearly in $F$. The bins are spaced linearly between a configured minimum and maximum simulation frequency.

For each reflection or diffraction point, the magnitudes $M_{\bullet}(f)$ are linearly interpolated from these $F$ bins onto a local frequency grid of $W$ samples at the client's sample rate $f_s$, with zero fill outside the simulated band. An inverse fast Fourier transform of this magnitude-only spectrum yields a symmetric, zero-phase kernel of length $W$, tapered with a Gaussian window of shape parameter $\kappa$ to limit its temporal extent. We place this kernel in the final synthetic signal sample output buffer at the sample nearest its propagation delay $r/c$ and sum across all points, so arrival times are accurate to within half a sample period rather than continuous.

In the original SonoTraceLab system, an additional post-processing step modeled, for example, the biological system's reception and emission organs. For humans, this is called the Head-Related Transfer Function (HRTF); for echolocating species, it is called the Echolocation-Related Transfer Function (ERTF), which is the product of the emission pattern and the HRTF. We made no modifications to this method, and it is available in the third-party API client for post-processing. For a more detailed description of the implementation of these filters, we refer to the original publications \cite{moto:c:irua:100095_stec_nove, jansenSonoTraceLabAraytracingBasedAcoustic2024}.

Without this post-processing step, assuming a signal $s_e(t)$ is emitted by the source $s$, we can calculate the final signal $y_m(t)$ arriving at receiver $m$ through convolution as follows:
\begin{equation}
\label{eq:convolve}
y_m(t) = h_{\text{tot},s,m}(t) * s_e(t)
\end{equation}

\section{EXPERIMENTAL RESULTS AND VALIDATION}
\label{sec:results}

The original work of SonoTraceLab was numerically validated in several bioacoustic scenarios \cite{jansenSonoTraceLabAraytracingBasedAcoustic2024}. In this section, we will focus the experiments on the new aspects and changes introduced into SonoTraceUE and provide example practical applications of the simulation framework, with a focus on supporting larger scenes with more dynamic applications of human-made acoustic systems for mobile robotics. Note that a comparative analysis with other simulation frameworks was outside the scope of this work due to the unique positioning of the proposed system. Existing commercial solvers are not always readily available for purchase. Furthermore, the absence of standardized formats for input configurations and output results precludes reliable, reproducible benchmarking of specific experimental scenarios across disparate software ecosystems. All experiments were run on a laptop with an NVIDIA RTX 3080 Ti Laptop GPU (\SI{16}{\giga\byte}) and an Intel i9-12900HK CPU with \SI{32}{\giga\byte} of memory. 

Table~\ref{tab:parameters} lists the full set of simulation parameters used in the two application experiments. These parameters are of two kinds. Some are fixed by physical or logical constraints of the experiment (e.g., the speed of sound, the simulated frequency band, the array sizes). Others (e.g., ray and bounce counts, curvature and area thresholds, material BRDF properties) were set empirically through testing rather than derived from measurement. In particular, although the framework supports full per-object BRDF definitions, we currently lack measured acoustic material properties for the surfaces involved, and the values used here should be read as reasonable working parameters rather than validated material data.

\begin{table}[!ht]
\caption{Simulation parameters used for the experiments of Section~\ref{sec:results}.}
\label{tab:parameters}
\centering
\footnotesize
\setlength{\tabcolsep}{2pt}
\begin{tabular}{p{0.30\linewidth}p{0.11\linewidth}p{0.25\linewidth}p{0.25\linewidth}}
\hline
\textbf{Parameter} & \textbf{Symbol} & \textbf{Section IV-A} & \textbf{Section IV-B}\\
\hline
\multicolumn{4}{l}{\textit{Curvature, Eq.~\eqref{eq:curvature}}}\\
Lower breakpoint & $A_1$ & \multicolumn{2}{c}{\SI{0.5}{\centi\meter\squared}}\\
Upper breakpoint & $A_2$ & \multicolumn{2}{c}{\SI{5}{\centi\meter\squared}}\\
Rejection breakpoint & $A_3$ & \multicolumn{2}{c}{\SI{200}{\centi\meter\squared}}\\
Minimum weight & $w_{\min}$ & \multicolumn{2}{c}{0.02}\\
Maximum weight & $w_{\max}$ & \multicolumn{2}{c}{2}\\
Curvature scale & $\eta$ & \multicolumn{2}{c}{1}\\
\hline
\multicolumn{4}{l}{\textit{Materials, Eq.~\eqref{eq:sigmoid}--\eqref{eq:brdfmap}}}\\
Smooth opening angle & $\varphi_{0}$ & \multicolumn{2}{c}{\SI{8.0}{\degree}--\SI{5.0}{\degree}}\\
Rough opening angle & $\varphi_{1}$ & \multicolumn{2}{c}{\SI{70.0}{\degree}}\\
Smooth reflection & $\rho_{0}$ & \multicolumn{2}{c}{10.0--8.0}\\
Rough reflection & $\rho_{1}$ & \multicolumn{2}{c}{0.025--0.05}\\
Slope ($\beta$, $k$) & $\sigma$ & \multicolumn{2}{c}{2.0}\\
Center ($\beta$, $k$) & $\mu$ & \multicolumn{2}{c}{0.4}\\
\hline
\multicolumn{4}{l}{\textit{Specular component}}\\
Initial rays & $N_{ray}$ & \num{600000} & \num{80000}\\
Maximum bounces & -- & 2 & 2\\
Maximum path length & -- & \SI{5}{\meter} & \SI{8}{\meter}\\
\hline
\multicolumn{4}{l}{\textit{Diffraction component}}\\
Candidates per mesh & $O_{m}$ & \num{8000} & \num{500}\\
\hline
\multicolumn{4}{l}{\textit{Sensing system}}\\
Speed of sound & $c$ & \SI{343}{\meter\per\second} & \SI{343}{\meter\per\second}\\
Emitters & $S$ & 1 & 1\\
Receivers & $R$ & 64 & 32\\
Frequency bins & $F$ & 14 & 14\\
Simulated band & -- & \num{20}--\SI{140}{\kilo\hertz} & \num{20}--\SI{85}{\kilo\hertz}\\
Emitted signal & $s_e(t)$ & recorded call & \SI{2.5}{\ms} chirp\newline{[}\num{25}--\SI{50}{\kilo\hertz}{]}\\
\hline
\multicolumn{4}{l}{\textit{Signal synthesis (MATLAB API client)}}\\
Sample rate & $f_s$ & \multicolumn{2}{c}{\SI{450}{\kilo\hertz}}\\
IR filter kernel length & $W$ & \multicolumn{2}{c}{512 samples}\\
Kernel window shape & $\kappa$ & \multicolumn{2}{c}{5}\\
Output IR length & -- & \multicolumn{2}{c}{14000 samples}\\
\end{tabular}
\end{table}

\begin{figure*}
    \centering
    \includegraphics[width=0.98\linewidth]{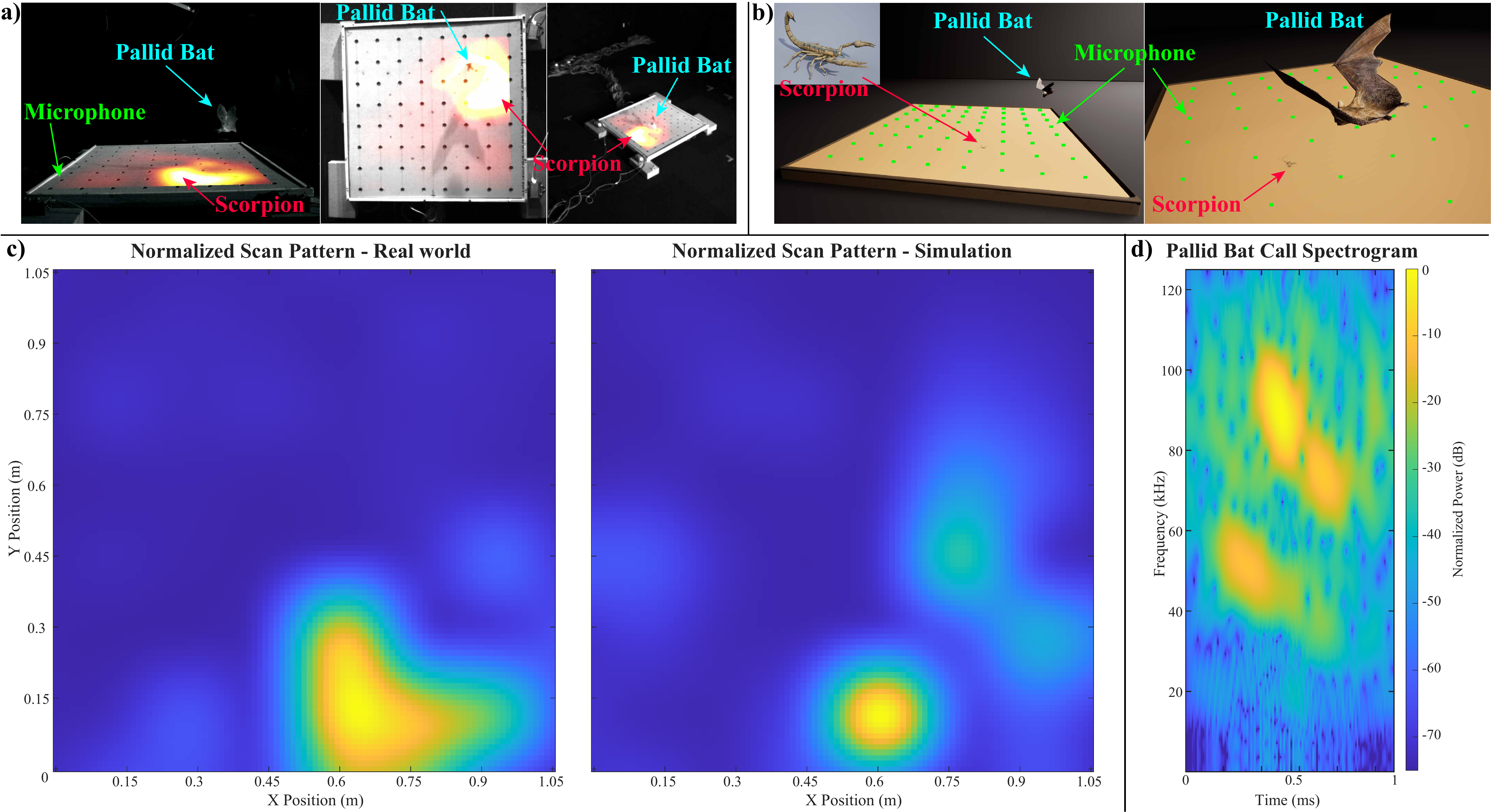}
    \caption{Comparison of the real-world experimental data of a single frame and its simulated recreation of the pallid bat hunting behavior. a) Three camera views captured simultaneously of the bat approaching its scorpion prey in the real-world experiment. This scorpion is sitting on a flat surface, with the 64-microphone array embedded within it. The acoustic intensity of the echolocation (scan pattern) is shown in a color gradation overlay. b) The same frame recreated with SonoTraceUE in Unreal Engine with the same poses for the microphones, scorpion, and bat. The locations of all 64 microphones are shown with green dots. A close-up of the 3D model used for the scorpion is shown in the top-left corner. c) The scan patterns, each normalized to its own maximum, showing the relative acoustic intensity across the array. They show the attention zone of the bat's emitted call. The comparison is spatial only and contains no timing information: the location of the attention zone is the same for the real-world and simulated results, while the shape of the pattern differs because the pallid bat's emission directivity is not modeled. d) The spectrogram of the echolocation signals during this measurement frame of the approaching bat, used by the simulation.}
    \label{fig:passive_experiment}
\end{figure*}

\subsection{BIOACOUSTICS}
\label{subsec:bioacoustics_experiments}

In a first experiment, we used SonoTraceUE to simulate a biological acoustic system, specifically pallid bats (\textit{Antrozous pallidus}). Verreycken et al. performed a study of their hunting behavior where they were trained to hunt for a scorpion (\textit{Hadrurus arizonensis}) that was placed on a random position within a planar microphone array with 64 microphones \cite{moto:c:irua:182944_verr_trac}. This array could record the bats' echolocation call sequences while they searched, approached, and eventually caught the scorpion. By analyzing the bat call's acoustic intensity at each microphone, one can see the bat's attention point and its shape. We recreated this scenario in simulation using the same recorded call as the emitted signal within a specific measurement frame. The microphone array locations and the 3D position and rotation of the bat were recreated from the study data, including the bat's flight path data. In this scenario, we used only the passive component of the simulation. Figure \ref{fig:passive_experiment} shows the results. As shown in the acoustic intensity scan-pattern heat maps, the attention location matches that of the study. Still, the specific shape is not reproduced because the pallid bat's physical emission pattern was not modeled or simulated. Figure~\ref{fig:passive_experiment}c compares normalized spatial intensity patterns across the array, and the acoustic attention point falls at the same position in simulation and in the recording, validating the spatial consistency of the passive component. The comparison is spatial only, as the available recordings were analyzed per measurement frame. The ability to easily use external study data in a visual 3D editor, such as the exact flight paths, represents a significant workflow improvement. By enabling the simulation of passive-listening scenarios, SonoTraceUE expands the scope of potential research beyond active echolocation to include passive sensing, capabilities previously unsupported.

In a second experiment, we tested the framework's ability to allow further insight into how these species use echolocation in their natural habitats. In the same study by Verreycken et al., additional measurements were conducted in which the surface supporting the microphone array was varied between a wooden flat surface and a layer of pebbles, the latter representing a rough surface. Subsequently, they compared the level of focus during the call for both surface types and analyzed echolocation precision. The degree of sonar beam focus was higher, and the distribution of centroid distances to the scorpion prey was shorter for the flat wooden surface. 

\begin{figure*}[!t]
    \centering
    \includegraphics[width=1\linewidth]{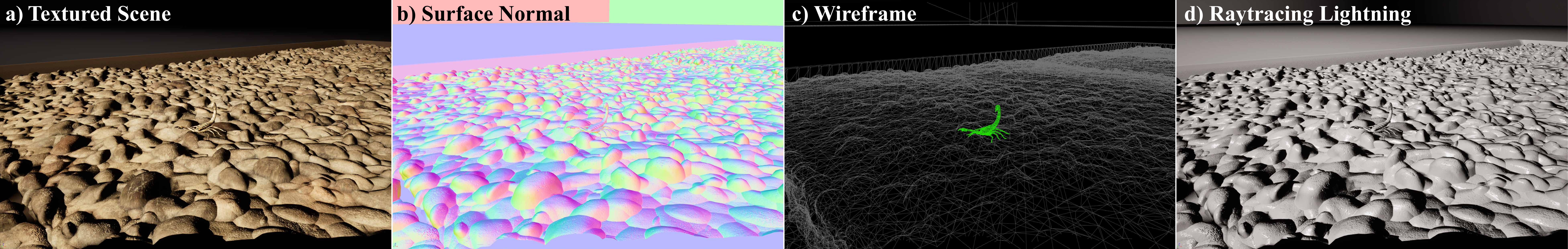}
    \caption{Various visualizations of the 3D scene from the same point of view used during the bioacoustic surface analysis experiment in simulation. The experiment shows how a rough surface affects the acoustic signature of a scorpion during bat echolocation hunting. a) A close-up of the 3D scene as created in Unreal Engine. b) The surface normal is visualized, which is used by the proposed method for calculating the curvature of the 3D meshes. c) The wireframe of the 3D meshes used in the scene showing the triangle-based geometry with the scorpion highlighted in green.  d) Only the light rendering result is shown here, where Unreal Engine also uses hardware-accelerated ray tracing to simulate light bouncing throughout the scene.}
    \label{fig:active_scene}
\end{figure*}

This supported the assumed theory that the prey camouflage themselves by living on rougher surfaces, such as the pebbled surface \cite{clareAcousticShadowsHelp2015a}. SonotraceUE now allows us to simulate and numerically determine the effect of the surface on the scorpion's acoustic signature. This scenario was recreated using 3D-modeled pebbles\cite{pebble3dmodel} and can be seen in Figure \ref{fig:active_scene}. 

We used the same position for the pallid bat as in the previous experiment, and the same call was recorded by the surface microphones. We simulated both the specular and diffraction components in this experiment using \SI{600000} initial rays that could bounce at most 2 additional times, and set the number of frequency bins to 14. To simulate these four scenarios, on average, the geometric pre-processing took \SI{270.54}{\ms} while the simulation of the specular and diffraction components took \SI{471.28}{\ms} and \SI{1025.13}{\ms} respectively.

\newpage 

\begin{figure*}[!ht]
    \centering
    \includegraphics[width=0.9\linewidth]{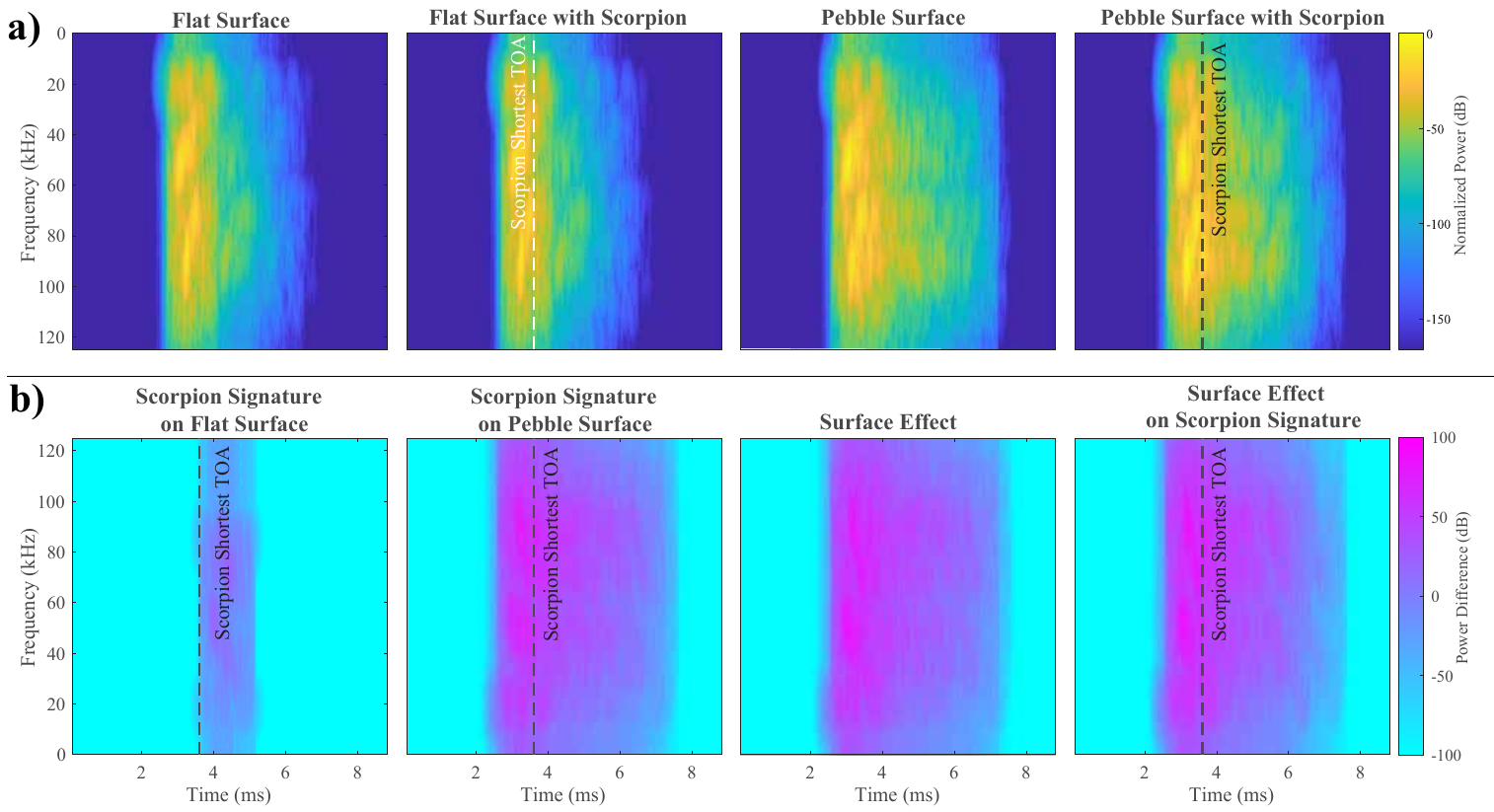}
    \caption{Spectrogram results of the bioacoustic surface analysis experiment in simulation to show the surface effect on the scorpion's acoustic signature when it is hunted by a bat that uses echolocation. We also add a vertical line marking the first arrival of the echo that reflects on the scorpion itself when it was present in the scenario.  a) These baseline spectrograms show the four distinct scenarios that were simulated from left to right: a flat surface baseline, a flat surface with target scorpion, a pebble surface baseline, and a pebble surface with target scorpion. All four panels are levels in \si{\decibel} referred to one common reference, the global maximum across the four scenarios, so that they are mutually comparable.  b) Difference panels obtained by coherently subtracting the complex spectra of two scenarios and converting the magnitude of that difference to \si{\decibel}, isolating the acoustic contribution added or removed between them. These use their own fixed \SI{-100}{}--\SI{100}{\decibel} color scale, not the shared reference of a). From left to right: the scorpion's acoustic signature (flat w. scorpion - flat), the scorpion's acoustic signature as influenced by specular reflections on the pebbles (pebbles w. scorpion - pebbles), the surface effect of the pebbles (pebbles - flat), the surface effect of the pebbles when scorpion is present (pebbles w. scorpion - flat w. scorpion).}
    \label{fig:active_experiment}
\end{figure*}

Figure \ref{fig:active_experiment}a shows the spectrogram results of this experiment, where we are varying the surface (flat vs. pebbles) and the presence of the target (scorpion vs. no scorpion). This arrangement yields four distinct measurement scenarios: (1) a flat surface baseline, (2) a flat surface with target scorpion, (3) a pebble surface baseline, and (4) a pebble surface with target scorpion. All four panels of Figure \ref{fig:active_experiment}a are expressed as levels in \si{\decibel} with respect to a single common reference, namely the global maximum across the four scenarios, so that the panels are directly comparable to one another.

\newpage

Furthermore, to gain further insight into the pebbles' surface effects and acoustic signatures, we created difference spectrograms for these scenarios, shown in Figure~\ref{fig:active_experiment}b, which isolate the acoustic contribution added or removed between two scenarios. The panels show that on the flat surface the prey introduces a compact, well-localized change in level, whereas on the pebbled surface the same change is spread across time and frequency and is comparable in magnitude to the change produced by the surface alone. Within the stated approximations (a geometric-acoustic abstraction of the Helmholtz equation, a BRDF-based scattering model, and an unmodeled emission directivity), this is consistent with the acoustic-camouflage hypothesis of \cite{clareAcousticShadowsHelp2015a}. It is a plausible mechanism by which surface roughness masks the prey's signature, presented here as a simulation-supported hypothesis. This simulation similarly suggests, as a hypothesis, that the bat can remain successful on the rougher surface by making multiple passes from different angles.

This level of environmental fidelity highlights SonoTraceUE's critical advantage in handling scene complexity and simulating the diffraction and scattering effects of hundreds of pebbles in less than \SI{2}{\s}. With the proposed system, we can achieve large scenes with the mesh density required to resolve our scene's Helmholtz number, ensuring that sub-centimeter environmental features are treated as acoustically significant rather than mere surface roughness.

\subsection{ROBOTIC ACOUSTIC SENSING}
\label{subsec:robotics_experiments}
In a second series of experiments, we tested human-made acoustic systems to validate the usage of the proposed framework as a development and testing environment for robotic applications of acoustic sensing. We recorded an indoor dataset using a 3D sonar sensor \cite{laurijssen2025ruggedizedultrasoundsensingharsh}, an RGB camera, and a high-resolution LiDAR sensor mounted on a small mobile platform. The scene was an office lobby with objects such as benches, vending machines, radiators, trash bins, and (elevator) doors. 

The goal was to use the acoustic sensor to create the mobile platform's trajectory. The camera and LiDAR sensor served as high-resolution ground truth sensors. We created this trajectory in two steps: first, the system used acoustic odometry to estimate the motion of a mobile agent by decomposing the mobile platform's speed into linear and rotational components \cite{6331017}. In a second step, we ran BatSLAM, a bio-inspired Simultaneous Localization and Mapping (SLAM) algorithm where we use 3D acoustic images from the sonar sensor to estimate the position of the robot \cite{moto:c:irua:105051_stec_bats, moto:c:irua:133180_stec_spat}. The correlation between acoustic images is continuously calculated to perform loop closure detection, which indicates recognition of previously visited places and helps minimize error in the trajectory mapping, as in acoustic odometry. 

\newpage

\begin{figure}[!t]
    \centering
    \includegraphics[width=1\linewidth]{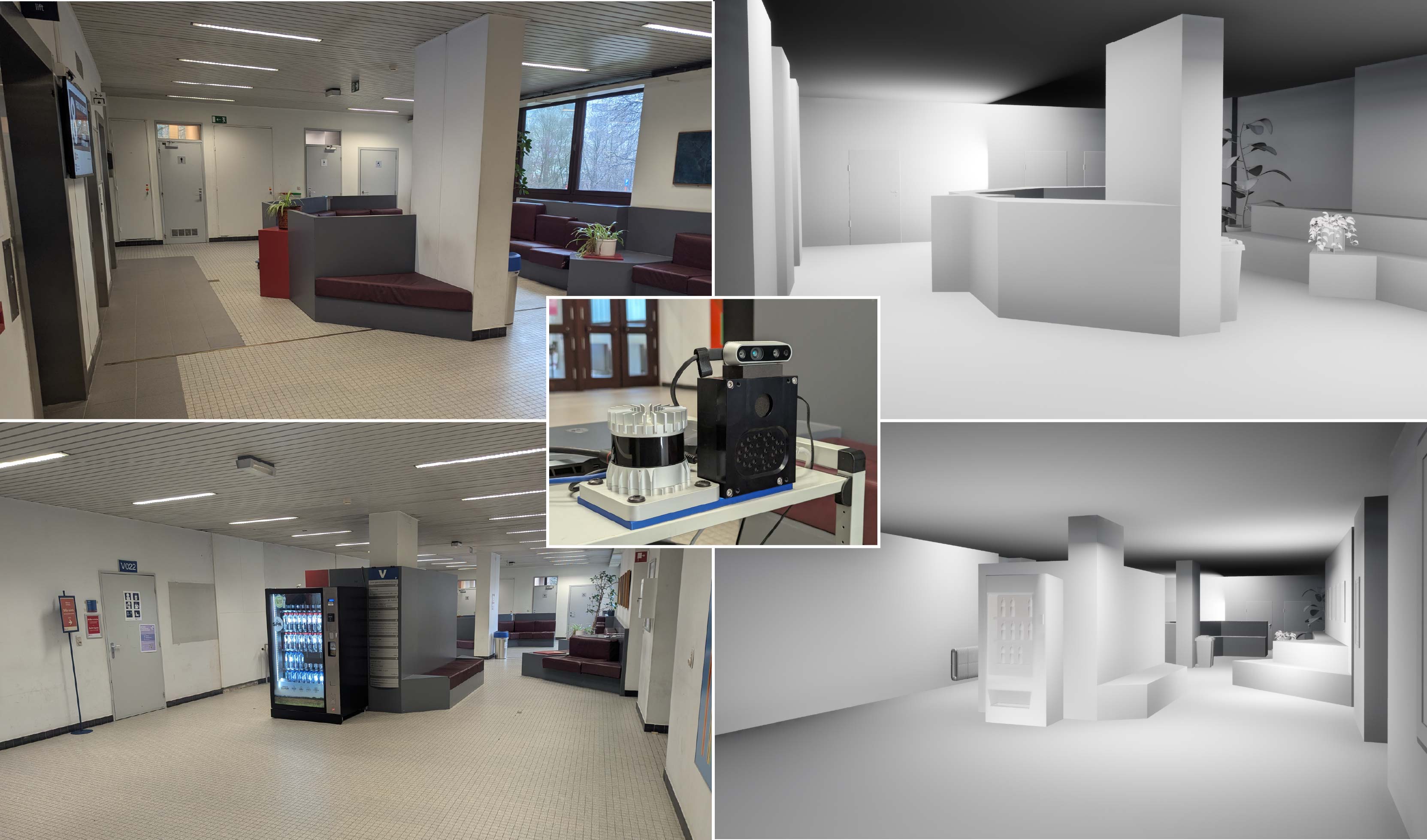}
    \caption{The robotics experiment within an indoor office lobby shown with real photos on the left and the virtual simplified recreation in Unreal Engine from the same point of view on the right. The photo in the middle shows the sensor configuration used for generating the dataset, including the 3D imaging sonar sensor \cite{laurijssen2025ruggedizedultrasoundsensingharsh}, the Ouster OS0-128 LiDAR, and the Intel RealSense D435i camera.}
    \label{fig:vbuilding}
\end{figure}

The goal of these experiments is to repeat them in a simplified virtual 3D recreation of this real environment, using the same acoustic sensor simulated with the proposed method, and to compare the acoustic odometry and SLAM results. The real-world scene and its virtual counterpart are shown in Figure \ref{fig:vbuilding} along with the sensor setup. The overall shape and layout of the real office lobby were recreated from 3D-scanned LiDAR data, with high-definition 3D models placed for several objects, as mentioned earlier. The hypothesis is that this simplified building geometry, combined with the unique acoustic signatures of the various high-resolution objects, will yield algorithm behavior comparable to that observed with the real sensor. 

The LiDAR sensor serves as the ground truth reference and was used as input to PIN-SLAM\cite{pinslam}, a state-of-the-art LiDAR SLAM algorithm, to generate a trajectory map. The simulated sensor is placed at the pose derived from the LiDAR trajectory, and the acoustic odometry and BatSLAM estimates are computed from the resulting signals but are never fed back into the motion of the platform. 
Any error in this LiDAR-derived trajectory is itself unknown, since no independent, higher-accuracy reference is available. Such error would introduce a gap between the pose the real sensor actually occupied during recording and the pose the simulated sensor is placed at, and could therefore contribute to the discrepancies reported below. This experiment design makes the real and simulated acoustic frames generated at supposedly identical viewpoints, since both are indexed to the same LiDAR-derived trajectory. 

The comparison isolates the sensor and environment combined response at a given pose rather than testing whether the two domains would follow the same path under their own motion. The acoustic system consists of a single emitter transmitting a \SI{2.5}{\ms} broadband chirp call between \SI{25}{\kHz} and \SI{50}{\kHz}, and a 32-element planar microphone array. Both the specular and diffraction components in this experiment were enabled with \SI{80000} initial rays that could bounce at most 2 additional times, and the number of frequency bins was set to 14. The offline computational cost of the geometric pre-processing was \SI{300}{\ms}. At the same time, during a single simulation frame, the specular and diffraction components took \SI{516.26}{\ms} and \SI{874.21}{\ms}, respectively, on average. These two stages together account for \SI{1.39}{\s} of GPU and CPU time per simulated frame, or at most \SI{0.72}{} simulated frames per second for the simulation stage alone, before any additional processing. This paper reports the performance of the geometric simulation only, the additional latency introduced by the API data transfer and the client-side signal synthesis of Subsection~\ref{subsec:signals} depends on other factors and the third-party client used and falls outside the scope of this work, so the true end-to-end cycle time is higher.

For creating the acoustic odometry results, we follow the same methodology as detailed in \cite{6331017}. Both simulated and real acoustic signals were converted into 2D acoustic images representing the horizontal plane in front of the mobile platform, with a resolution of \SI{0.5}{\degree} spanning \SI{-80}{\degree} to \SI{80}{\degree}. 

\begin{figure*}[!t]
    \centering
    \includegraphics[width=0.95\linewidth]{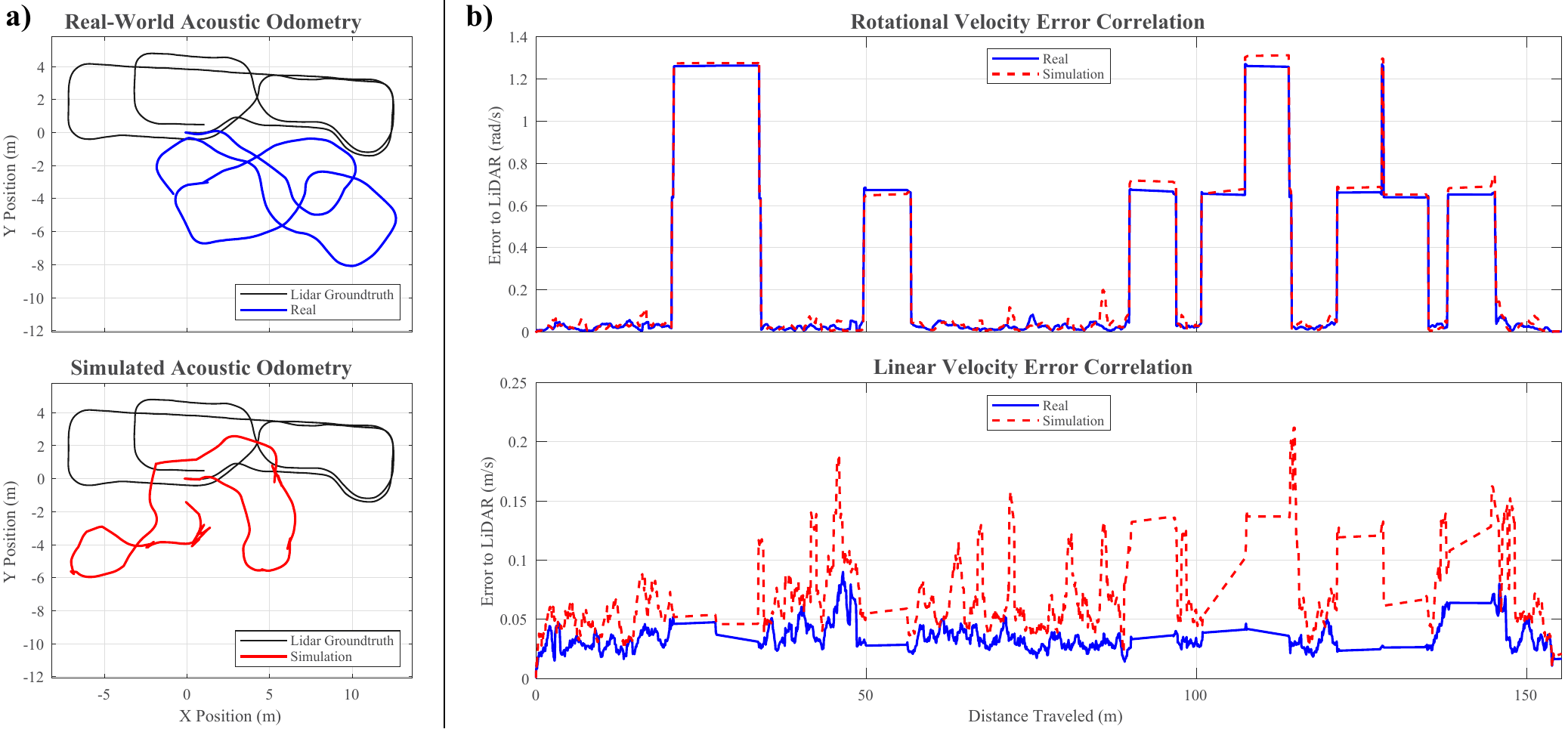}
    \caption{Results of the acoustic odometry for the robotics experiment. a) The trajectories of the real and simulated 3D sonar sensors compared to the ground truth as generated by a LiDAR sensor. The resulting acoustic odometry trajectory differs clearly. b) The per-frame absolute error of the rotational and linear velocity estimated by the acoustic odometry with respect to the LiDAR ground truth, for the real and for the simulated sensor. The two error traces agree closely for the rotational velocity and more loosely for the linear velocity, and peak in the same regions of the scene.}
    \label{fig:acoustic_odom_exp}
\end{figure*}

\newpage
The results of the acoustic odometry trajectories can be seen in Figure \ref{fig:acoustic_odom_exp}a. Both the real and simulated acoustic sensors exhibit drift relative to the LiDAR ground truth in both straight lines and corners. The simulation recreation does not perfectly mirror the stochastic drift path of the real data. To assess whether the proposed framework nevertheless reproduces the behavior of such acoustic systems and their applications, we compared the velocity estimation errors of the two domains, as shown in Figure \ref{fig:acoustic_odom_exp}b.

Figure \ref{fig:acoustic_odom_exp}b shows, for each frame, the absolute velocity error of the acoustic estimate with respect to the LiDAR ground truth, for the real and for the simulated sensor. The two error traces track each other closely for rotational velocity and more loosely for linear velocity, and both peak in the same geometric regions of the scene. The mean absolute rotational velocity error difference between the real and simulated acoustic odometry result was \SI{0.024}{\radian\per\second} with a standard deviation of \SI{0.059}{\radian\per\second}, while the same for the linear velocity resulted in a mean of \SI{0.037}{\meter\per\second} with a standard deviation of \SI{0.076}{\meter\per\second}. We believe the larger difference in linear velocity is because more reflections are present in the real-world due to surface imperfections that we did not model in our simplified virtual re-creation. The simulated odometry fails in the same places, in the same direction, and to a similar degree as the real one. It does not reproduce the trajectory itself, as Figure \ref{fig:acoustic_odom_exp}a makes clear. Agreement in failure locations is the property that matters for developing and stress-testing applications such as acoustic odometry.

\begin{figure*}[!t]
    \centering
    \includegraphics[width=1\linewidth]{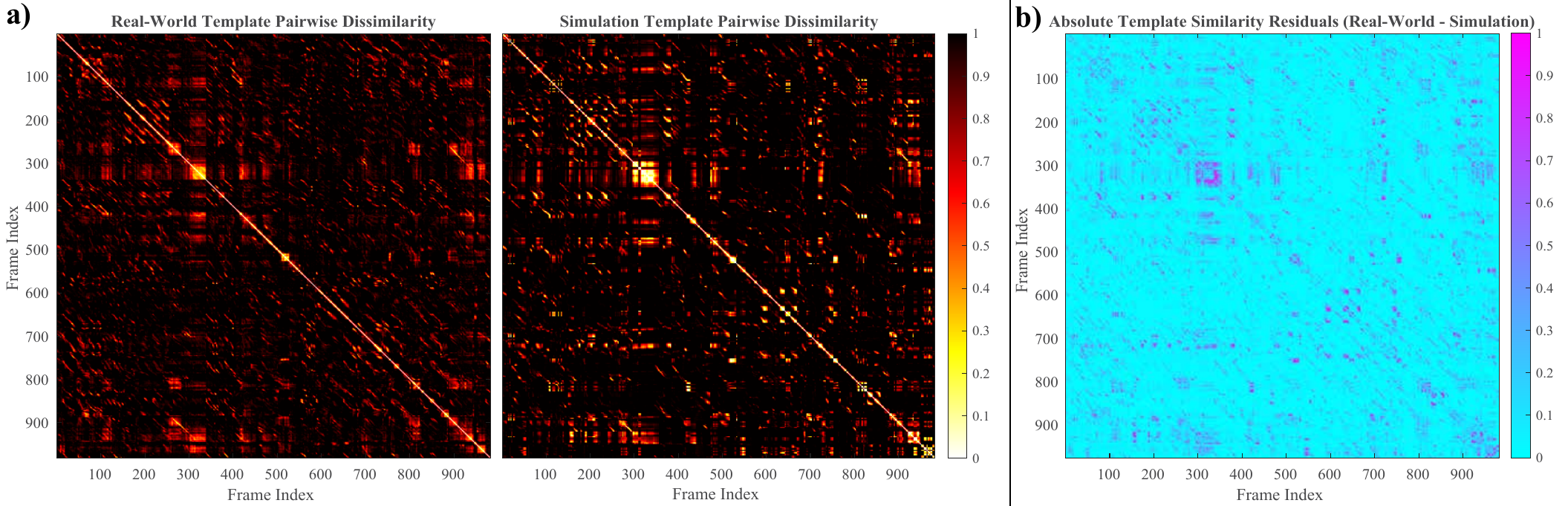}
    \caption{Results of the acoustic image template matching for the (Bat)SLAM application. a) Pairwise dissimilarity matrices between all frames along the full trajectory of the dataset, for the real 3D sonar acoustic images on the left and for the simulated sensor on the right. Each entry is the mean absolute difference between two acoustic images, so low (bright) values indicate high template similarity and mark loop-closure candidates. b) The absolute difference between the two matrices in a). The mean absolute residual between the real-world and simulated dissimilarity was 0.0745 with a standard deviation of 0.112.}
    \label{fig:template_matching}
\end{figure*}

To further assess the fidelity of the simulation, we evaluated the performance of the BatSLAM place recognition as described in \cite{moto:c:irua:105051_stec_bats} on the same datasets. Central to this bio-inspired SLAM algorithm is the ability to recognize previously visited locations by comparing the current acoustic image with a history of previous ones. We replicated this process by generating pairwise difference matrices for both the real and simulated trajectories, as shown in Figure \ref{fig:template_matching}a.

The two matrices share the same off-diagonal band structure, where trajectory revisits appear. Figure \ref{fig:template_matching}b shows the absolute difference between the real and simulated dissimilarity matrix, with a mean of 0.0745 and a standard deviation of 0.112 over the full trajectory on a $[0,1]$ scale. This indicates that the discrepancy between the two domains is small relative to the range on which the place-recognition threshold may operate.

\begin{figure}[!t]
    \centering
    \includegraphics[width=1\linewidth]{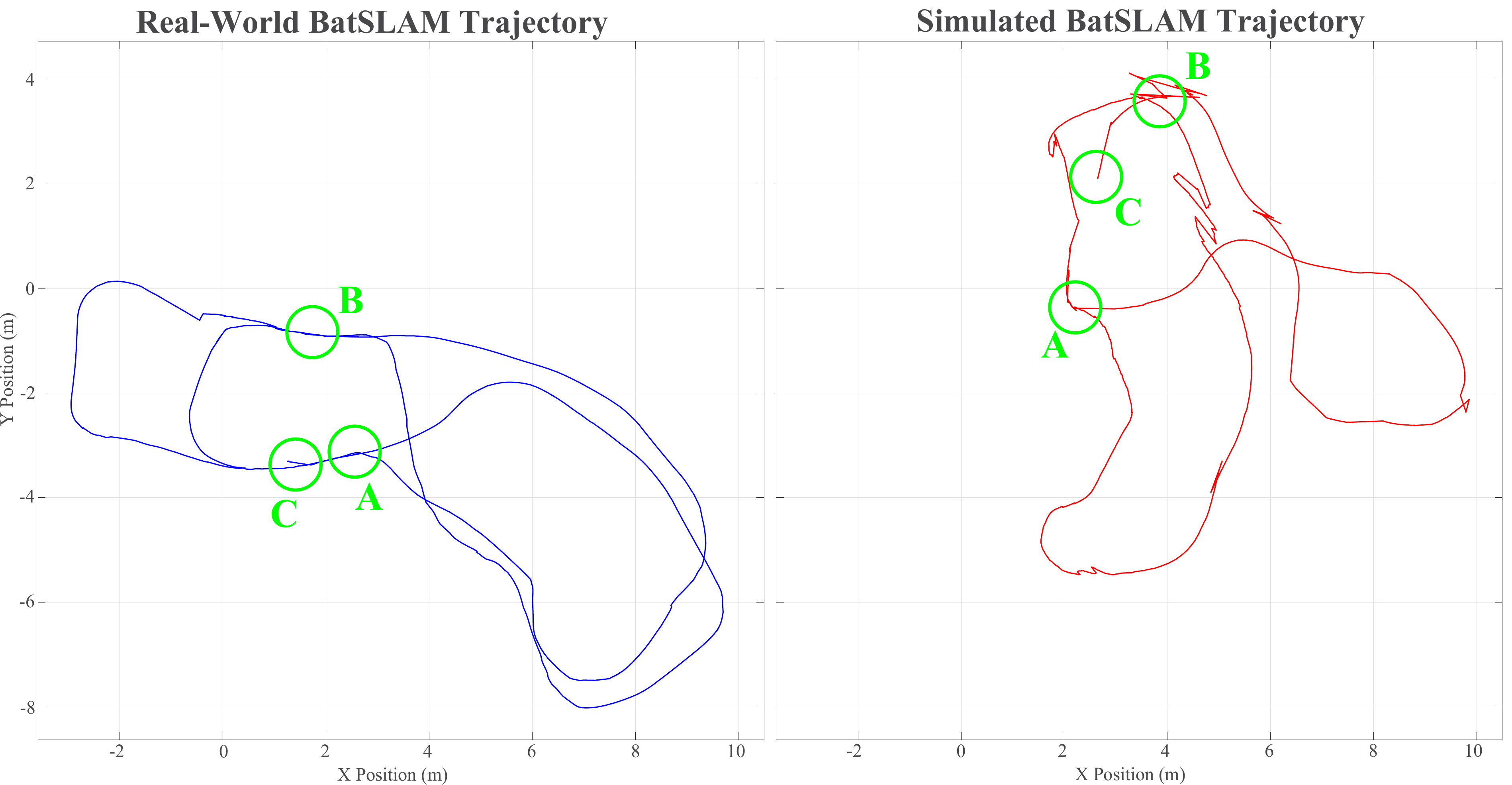}
    \caption{BatSLAM trajectory reconstruction integrating acoustic odometry with template-based loop closure detection. Three distinct loop closure zones are annotated as A, B, and C. a) The real-world trajectory successfully registers loop closures in all three zones, effectively mitigating cumulative odometry drift and preserving the global topological shape. b) The simulated trajectory successfully identifies loop closures at A and B but fails to register a detection at zone C, where the simulated dissimilarity does not fall below the acceptance threshold. The lack of correction at C results in uncorrected drift and significant deviation from the ground truth path topology. This case shows that agreement in the aggregate dissimilarity statistics of Figure~\ref{fig:template_matching} does not guarantee that every loop-closure candidate transfers from simulation to reality.}
    \label{fig:batslam}
\end{figure}

Applying the template matching within the full BatSLAM algorithm yields the trajectories of Figure \ref{fig:batslam}. The real dataset registers loop closures in all three annotated zones A, B and C, whereas the simulated dataset registers A and B but misses C. 

\newpage 
At that revisit, the simulated dissimilarity does not fall below the acceptance threshold of the place-recognition stage. We attribute this to the simplified virtual reconstruction of that part of the lobby, where fewer distinctive high-resolution objects were modeled than elsewhere along the trajectory, reducing the acoustic distinctiveness of that location compared to the real-world data, where more high-resolution information was present. The experiment supports the framework as a testbed for developing and stress-testing acoustic place recognition, with the level of geometric and material detail in the virtual reconstruction as the limiting factor.

\subsection{COMPUTATION ANALYSIS}
\label{subsec:computation_experiments}

To demonstrate the scalability and real-time capabilities of the proposed SonoTraceUE framework, we conducted a comprehensive computational performance analysis. The evaluation's primary objective was to quantify the system's resource utilization and execution latency across varying degrees of scene complexity and simulation fidelity. 

Specifically, we investigated the impact of increasing geometric density, measured in triangle count, on both memory footprint and compute time during the mesh preparation step. Furthermore, we analyze the runtime performance of the specular and diffraction components under increasing computational load, isolating the effects of increasing ray counts, receiver and emitter densities, the number of allowed bounces, and diffraction parameters. These metrics provide insight into the algorithmic efficiency of the hardware GPU-accelerated pipeline and establish the operational boundaries for real-time acoustic simulation in large-scale environments.

\begin{figure}[!b]
    \centering
    \includegraphics[width=1\linewidth]{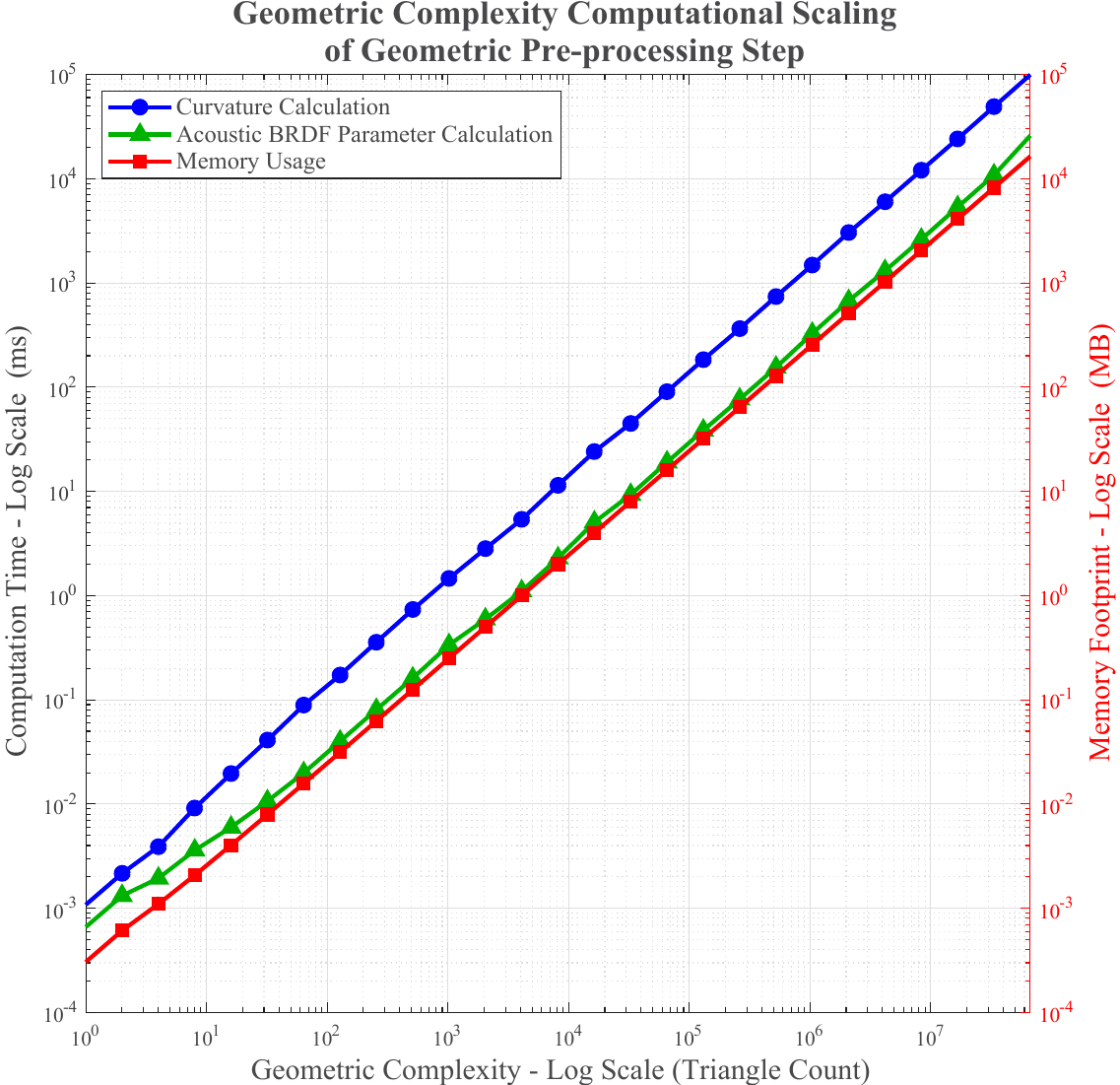}
    \caption{Computational scalability of the offline pre-processing stage as a function of geometric complexity defined by the triangle count. The graph decomposes the processing latency into purely geometric curvature estimation and frequency-dependent acoustic BRDF parameter mapping, and also shows the memory footprint. The data represent the mean of five execution runs with a fixed high-fidelity spectral resolution of 20 frequency bins. The results confirm linear scalability for both compute time and memory.}
    \label{fig:comp_offline_triangle}
\end{figure}

\begin{figure}[!t]
    \centering
    \includegraphics[width=0.95\linewidth]{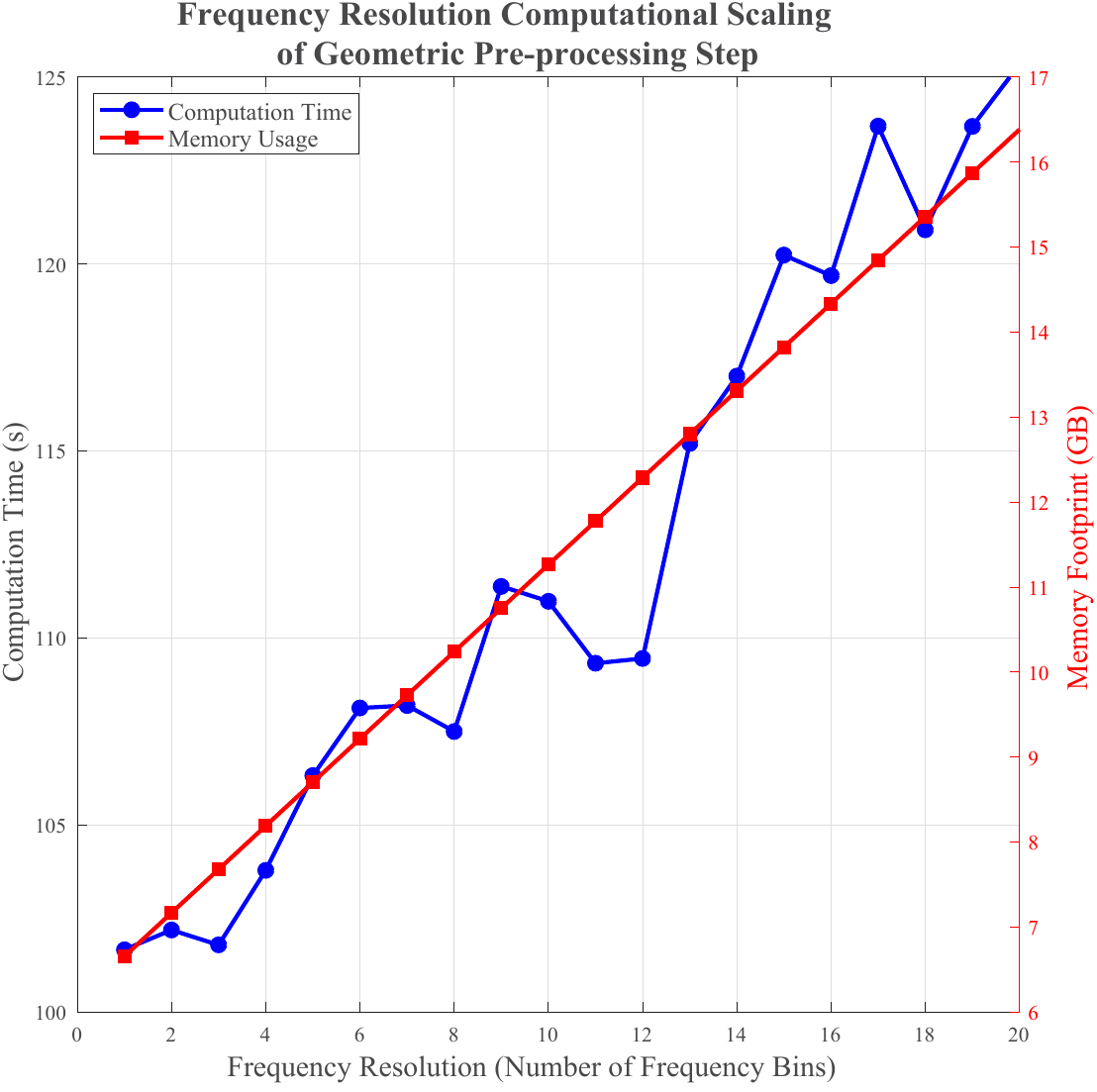}
    \caption{Impact of the spectral resolution defined by the number of frequency bins on resource utilization for a fixed high-density mesh of \si[parse-numbers=false]{2^{26}} triangles. This demonstrates that increasing simulation fidelity, by increasing frequency resolution, imposes a proportional linear cost on pre-processing time and memory requirements.}
    \label{fig:comp_offline_freq}
\end{figure}

For a first analysis, we looked at the first geometric pre-processing step, which happens offline before the simulation starts. It calculates the curvature for each triangle of the 3D scene as well as the BRDF properties for each triangle and frequency bin combination as detailed in Subsection \ref{subsec:pre-processing}. The memory footprint of the data stored in memory is equal to $128 + 8 \cdot T(12 + F)$ bytes, where $T$ is the triangle count and $F$ is the frequency bin count. In this experiment, we scaled the triangle count in powers of two from 1 to \si[parse-numbers=false]{2^{26}} and the number of frequency bins from 1 to 20, and ran five runs with all combinations of triangle count and frequency resolution. Firstly, we examine computational scaling with increasing geometric complexity (increasing triangle count) for both computation time and memory usage when saving the calculated curvature and BRDF data. The mean across these five runs is shown in Figure \ref{fig:comp_offline_triangle} for a fixed frequency resolution of 20 bins, in a worst-case scenario. We observe a maximum memory usage of \SI{16384}{\mega\byte} and computational times of \SI{99.43}{\s} and \SI{25.91}{\s} for the curvature and BRDF parameter calculations, respectively. All aspects scale linearly, as expected. Secondly, the same analysis was done for scaling with the frequency bin resolution and is shown in Figure \ref{fig:comp_offline_freq} for a fixed geometric complexity resolution of \si[parse-numbers=false]{2^{26}} triangles, in a worst-case scenario, showing a similar linear scaling. Since this step is performed once before the simulation starts, these results indicate that even for very large, dense geometric scenes with high spectral resolution, the additional simulation resource utilization remains reasonable.

Following geometric pre-processing, the main simulation loop consists of three sequential stages: ray tracing shader execution, specular component simulation, and diffraction component simulation. We first analyze the execution time and GPU memory usage of the ray tracing shader. In this experiment, the number of initial rays was scaled in powers of two from 1 to \SI{2097152}(\si[parse-numbers=false]{2^{21}}), while the number of specular bounces varied between 1 and 10. Five runs were performed for every combination to capture the variance. To stress-test the system, the scene setup ensured that every ray completed the maximum number of bounces. 

\newpage 

\begin{figure}[!t]
    \centering
    \includegraphics[width=0.95\linewidth]{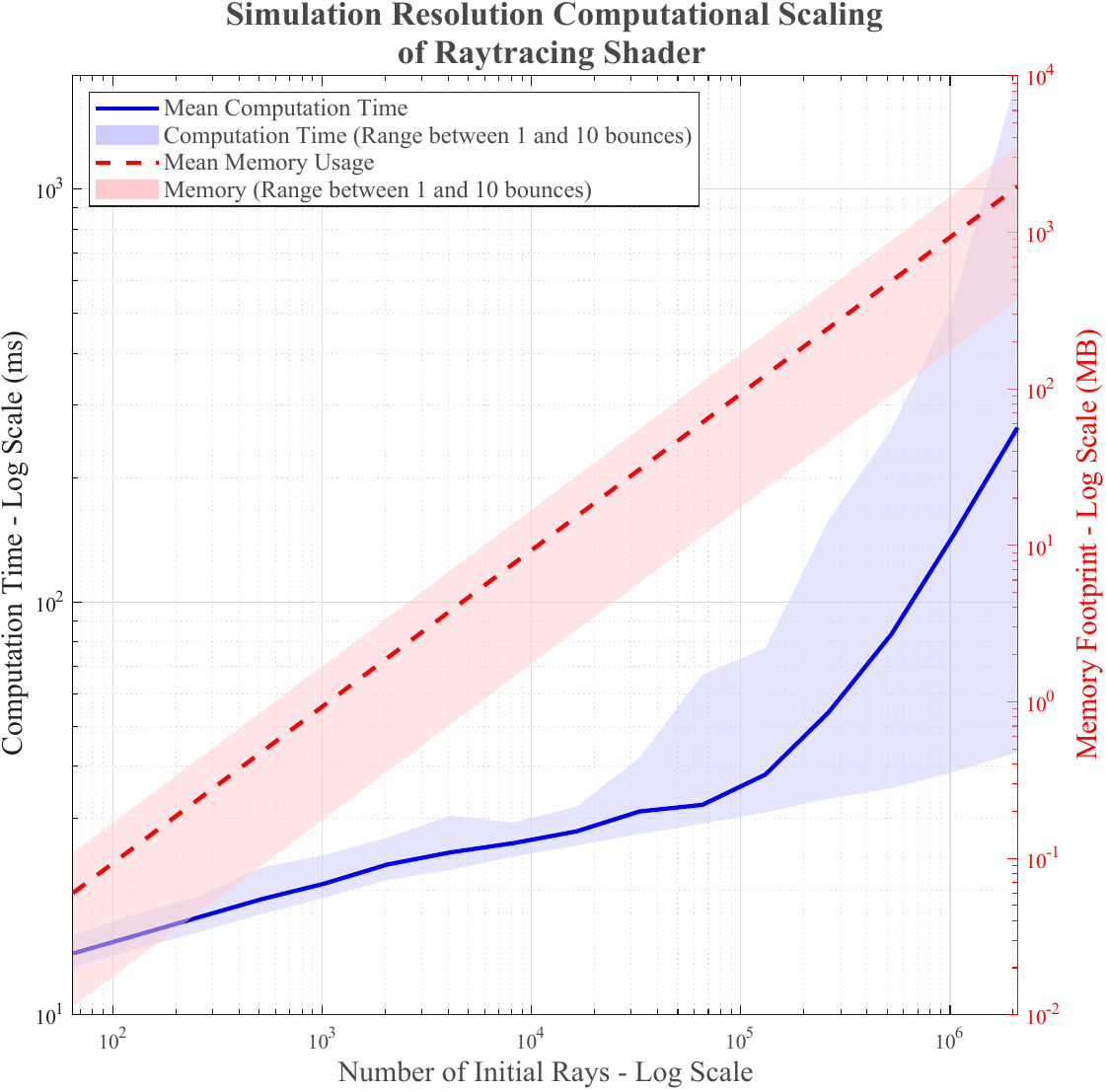}
    \caption{ray tracing shader performance scaling with initial ray count. The solid line represents the mean execution time across all bounce depths, while the shaded region indicates the range between the minimum (1 bounce) and maximum (10 bounces) workload. The secondary axis tracks the GPU memory utilization. The number of initial rays was scaled in powers of two from 1 to \si[parse-numbers=false]{2^{21}}.}
    \label{fig:comp_raytracing}
\end{figure}

Figure \ref{fig:comp_raytracing} illustrates the results, where the shaded region represents the variability between the minimum (1 bounce) and maximum (10 bounces) configurations, and the solid line indicates the mean. Memory usage scales linearly with the number of initial rays and bounces, as expected. Computation time follows a similar linear trend initially but exhibits exponential behavior beyond approximately \SI{65536} initial rays. The Unreal Engine and DXR documentation do not offer any further insight into why this occurs with these large ray counts. We speculate this may not happen on larger High-Performance Compute (HPC) platforms with more available GPU resources. For the maximum load of \si[parse-numbers=false]{2^{21}} initial rays with 10 bounces, we observed a computation time of \SI{1882.02}{\ms} and a memory usage of \SI{3536}{\mega\byte}. While useful for benchmarking, simulating this number of rays and specular reflection points exceeds the practical memory constraints of typical workstations when also simulating the specular magnitude of each point in the next step, restricting such scenarios to HPC environments.

\begin{figure*}[!t]
    \centering
    \includegraphics[width=0.9\linewidth]{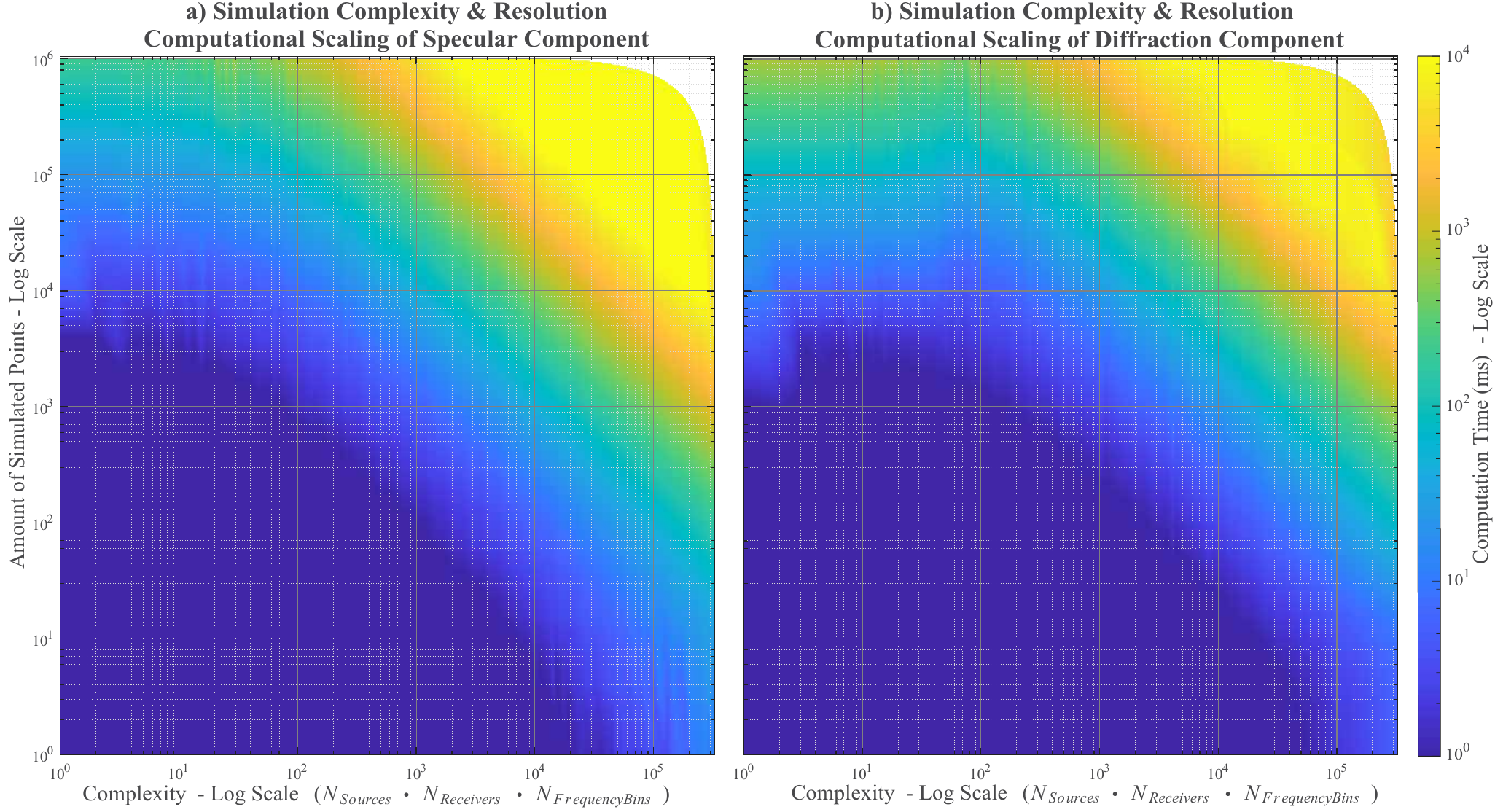}
    \caption{Computational performance landscape of the simulation components. The heatmaps illustrate the scaling of execution time with simulation complexity and the number of simulation points. a) Shows this computational landscape for the specular component. b) Shows this computational landscape for the diffraction component. }
    \label{fig:comp_simulation_time}
\end{figure*}

Upon completion of the ray tracing shader, the reflection points are parsed on the CPU to compute the specular and diffraction components. Beyond the raw point count, the computational load depends on several scaling parameters: the number of sources ($S$), the number of receivers ($R$), and the number of frequency bins ($F$). We define the product of these terms as the simulation complexity. Both specular and diffraction components scale with a complexity of $\mathcal{O}(N \cdot S \cdot R \cdot F)$, where $N$ represents the number of simulation points. Parallelization across $N$ threads reduces computation time proportionally to the number of available CPU cores. 

\begin{figure}[!t]
    \centering
    \includegraphics[width=1\linewidth]{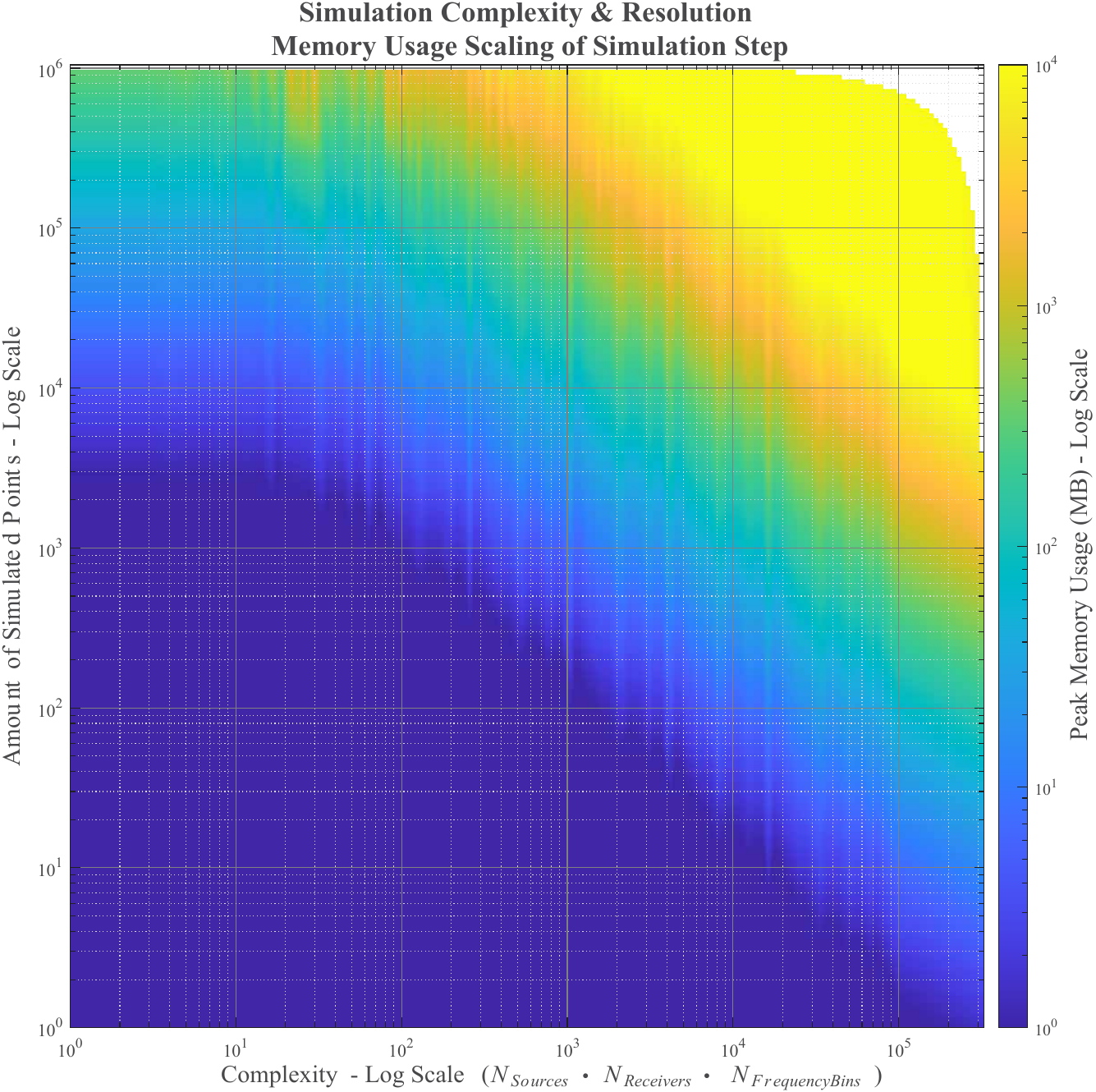}
    \caption{The memory footprint scaling for the CPU-based simulation components. The memory footprint for specular and diffraction is the same when using the same complexity input parameters. The heatmap visualizes memory requirements relative to simulation complexity and the number of active simulation points, highlighting the constraints in high-fidelity scenarios.}
    \label{fig:comp_simulation_memory}
\end{figure}

\newpage A comprehensive parameter sweep was conducted, with five runs per combination. The parameter ranges were constrained as follows: points from 1 to \SI{1048576}(\si[parse-numbers=false]{2^{20}}), sources from 1 to 32, receivers from 1 to 1024, and frequency bins from 1 to 20. Note that we did not run any combination that exceeded \SI{20}{\giga\byte} of memory due to limitations of the test system. Figure \ref{fig:comp_simulation_time}a presents the performance landscape for the specular component as an interpolated logarithmic plot, mapping average execution time against complexity and point count. Figure \ref{fig:comp_simulation_time}b illustrates the corresponding landscape for the diffraction component. 
The worst-case computation we found was \SI{32.88}{\s} with \si[parse-numbers=false]{2^{20}}, \si[parse-numbers=false]{2^{18}} points and a simulation complexity of \SI{13312}(16 emitters, 64 receivers, and 13 frequency bins).
Finally, Figure \ref{fig:comp_simulation_memory} details the peak memory usage, which is consistent across both specular and diffraction stages. The worst-case memory usage found was \SI{19585}{\mega\byte} with \si[parse-numbers=false]{2^{20}} points and a complexity of \SI{3200}(5 emitters, 64 receivers, and 10 frequency bins).

Together, these results show that achieving real-time performance that mimics real-world acoustic systems, both biological and human-made, depends heavily on configuration rather than being a fixed property of the framework. For the lighter configurations characterized above, with modest ray counts, moderate bounce depths, and few receivers, the pipeline reaches interactive rates suitable for iterative scenario design and closed-loop simulation. For the higher-fidelity robotics configuration used in Subsection~\ref{subsec:robotics_experiments}, the specular and diffraction stages alone already total \SI{1.39}{\s} per frame, before any additional client-side processing. Our computational analysis lets us locate any intended configuration on this trade-off and estimate its achievable rate accordingly.

\section{CODE AVAILABILITY AND USAGE}
\label{sec:availability}
We made the simulation framework open-source on GitHub \cite{sonotraceuegithub}. The source code is provided as a complete Unreal Engine project containing the SonoTraceUE plugin and a few example scenarios. These examples help users get started efficiently. As mentioned in Subsection \ref{subsec:Architecture}, the signal generation is available via an API in an external third-party client. It is currently implemented as both an open-source MATLAB Toolbox \cite{sonotraceuematlabtoolboxfilexchange} and a Python package\cite{sonotraceuepythonmodule}. All results reported in this paper were produced with SonoTraceUE release v3.1.0 on Unreal Engine v5.4, together with the MATLAB API client release 3.0.0\cite{sonotraceuematlabtoolboxfilexchange}. The workflow leverages the client-server architecture. First, the user opens the provided Unreal Project, where the 3D scene is defined. Unlike the previous MATLAB-only implementation, which loads meshes via a script, the environment is now configured directly within the Unreal Editor. This allows for visual placement of the sensor and emitter actors, as well as the assignment of acoustic material properties to objects using the editor's native interface. Furthermore, entire world simulations and dynamic scenarios can be scripted within the Unreal framework. Once the scene is prepared, the engine starts the simulation and initializes the API server. On the client side, the user utilizes the provided API client to interact with the running simulation. The user establishes a connection to the server to retrieve the simulation configuration. The user can trigger the simulation via the API and control the sensor's position or the position of its parent mobile system. The engine performs a hardware-accelerated simulation and returns the resulting point cloud containing all simulation data over the API connection, which can then be further processed.

\section{LIMITATIONS}
\label{sec:limitations}
While SonoTraceUE significantly advances the state of geometric acoustic simulation, it is not without limitations. First, as with all geometric approaches, it remains an approximation of the Helmholtz equation. While it offers a computationally tractable alternative to wave-based solvers (FEM/BEM) for high-frequency, large-scale environments, it cannot fully replace exact numerical methods for low-frequency or small-scale bounded problems. Second, the current diffraction model relies on a local curvature metric. This approach prioritizes computational efficiency over full principal curvature computation because Unreal Engine's mesh interface does not provide complete one-ring neighborhood information. Consequently, this approach requires meshes with adequate tessellation density at acoustically significant features to ensure that the per-triangle curvature variation accurately captures local geometry. Furthermore, because the curvature calculation is performed offline, animated skeletal meshes are not supported for the diffraction component calculation. We also see potential to extend the hardware-accelerated pipeline to further offload workloads to the GPU, for example, by integrating real-time geometric curvature analysis and the entire specular magnitude calculation within the compute shaders. More broadly, the framework's per-object BRDF model is only as useful as the material data available to populate it, and no systematic dataset of measured acoustic reflection properties exists for common natural or man-made surfaces across the ultrasonic band.

Several further limitations follow from Section~\ref{sec:methods} and from the scope of the experiments. Neither the specular nor the diffraction component is normalized by its sampling density, so synthesized levels are comparable only within a fixed configuration. The difference spectrograms used in Subsection~\ref{subsec:bioacoustics_experiments} are logarithmic ratios of coherent sums and cannot isolate the echo of an individual scatterer, so we state the conclusions drawn from them as hypotheses. The pallid bat's emission directivity is not modeled, limiting any comparison of beam shape to the location of the acoustic attention point. Finally, the robotic experiments replay a LiDAR-derived trajectory instead of closing the control loop on the acoustic estimate, and the achievable simulation rate for the configurations used there, below one frame per second end to end, is an order of magnitude short of the physical sensor's frame rate.

\section{CONCLUSION}
\label{sec:conclusion}
In this paper, we present SonoTraceUE, a novel high-fidelity acoustic simulation framework implemented as a plugin for Unreal Engine. Building upon the theoretical foundation of the previous work, SonoTraceLab, this new architecture transitions from a MATLAB-based solver to a hardware-accelerated, real-time 3D engine. By leveraging hardware-accelerated ray tracing for specular reflections while keeping the curvature-based Monte Carlo approximation for diffraction, SonoTraceUE addresses the critical scalability and latency limitations of its predecessor. \newpage 

We demonstrate the framework's validity through a series of bioacoustic and robotic experiments, showing that it reproduces relevant acoustic cues of biological echolocation and generates synthetic data on which robotic algorithms such as acoustic odometry and place recognition can be developed and stress-tested. Depending on the simulation configuration and computation hardware available, the architecture can also support closed-loop simulation. 

Looking ahead, several avenues for further development are planned to enhance the framework's fidelity and accessibility. Future iterations will expand the ray tracing shader to include acoustic transmission and refraction, enabling the simulation of multi-medium environments such as underwater acoustics or sound propagation through complex thin-walled structures. Additionally, to further reduce latency for real-time closed-loop applications, we aim to integrate accelerated C++ math and digital signal processing libraries, such as optimized fast Fourier transform implementations, directly within the plugin to minimize overhead during signal synthesis caused by the current API and reliance on a third-party client for signal generation. Additionally, we are exploring a porting of the framework to NVIDIA Isaac Sim and Cosys-AirSim, which would allow SonoTraceUE to serve as the acoustic sensing backend for a standardized, high-fidelity robotics simulation ecosystem. Beyond research, the visual nature of the Unreal Engine environment positions SonoTraceUE as a powerful educational tool. It allows students and researchers to visualize otherwise invisible acoustic phenomena, bridging the gap between theoretical acoustics and practical sensor behavior. 

\section*{DECLARATION OF GENERATIVE AI USAGE}
Generative AI was used in the creation of this article, specifically Microsoft's Copilot LLM, to improve the readability of some figures in this manuscript by generating MATLAB code to enhance their layout and style.

\bibliographystyle{IEEEtran}
\bibliography{main}
\end{document}